\documentclass{ieee-ojvt}
\usepackage{cite}
\usepackage{amsmath,amssymb,amsfonts}
\usepackage{graphicx,color}
\usepackage{textcomp}
\usepackage{psfrag}
\usepackage{subcaption} 
\usepackage{epsfig}
\usepackage{pifont}
\usepackage{array}
\usepackage[T1]{fontenc}
\usepackage{algorithm,algorithmic}
\usepackage{hhline} 
\usepackage{longtable}
\usepackage{diagbox}
\usepackage{tabularx}
\usepackage{comment}
\usepackage{caption}
\usepackage{multirow}
\usepackage{hyperref}
\usepackage{stfloats} 

\usepackage{acro}
\DeclareAcronym{isac}{
  short=ISAC,
  long=Integrated Sensing and Communication,
}
\DeclareAcronym{psk}{
  short=PSK,
  long=Phase Shift Keying,
}
\DeclareAcronym{bpsk}{
  short=BPSK,
  long=Binary Phase Shift Keying,
}
\DeclareAcronym{qpsk}{
  short=QPSK,
  long=Quadrature Phase Shift Keying,
}
\DeclareAcronym{qam}{
  short=QAM,
  long=Quadrature Amplitude Modulation,
}
\DeclareAcronym{ber}{
  short=BER,
  long=Bit Error Rate,
}
\DeclareAcronym{mmse}{
  short=MMSE,
  long=Minimum Mean Square Error,
}
\DeclareAcronym{rmse}{
  short=RMSE,
  long=Root Mean Square Error,
}
\DeclareAcronym{mse}{
  short=MSE,
  long=Mean Square Error,
}
\DeclareAcronym{dft}{
  short=DFT,
  long=Discrete Fourier Transform,
}
\DeclareAcronym{idft}{
  short=IDFT,
  long=Inverse Discrete Fourier Transform,
}
\DeclareAcronym{isdft}{
  short=ISDFT,
  long=Inverse Symplectic Discrete Fourier Transform,
}
\DeclareAcronym{sdft}{
  short=SDFT,
  long=Symplectic Discrete Fourier Transform,
}
\DeclareAcronym{dct}{
  short=DCT,
  long=Discrete Cosine Transform,
}
\DeclareAcronym{aoa}{
  short=AoA,
  long=Angle of Attack,
}
\DeclareAcronym{aod}{
  short=AoD,
  long=Angle of Departure,
}
\DeclareAcronym{los}{
  short=LoS,
  long=Line of Sight,
}
\DeclareAcronym{nlos}{
  short=NLoS,
  long=Non-Line of Sight,
}
\DeclareAcronym{ofdm}{
  short=OFDM,
  long=Orthogonal Frequency-Division Multiplexing,
}
\DeclareAcronym{ofdma}{
  short=OFDMA,
  long=Orthogonal Frequency-Division Multiple Access,
}
\DeclareAcronym{sc-fdma}{
  short=SC-FDMA,
  long=Single Carrier Frequency-Division Multiple Access,
}
\DeclareAcronym{im}{
  short=IM,
  long=Index Modulation,
}
\DeclareAcronym{im-ofdm}{
  short=IM-OFDM,
  long=\ac{im} \ac{ofdm},
}
\DeclareAcronym{otfs}{
  short=OTFS,
  long=Orthogonal Time Frequency Space,
}
\DeclareAcronym{otsm}{
  short=OTSM,
  long=Orthogonal Time Sequency Multiplexing,
}
\DeclareAcronym{im-otfs}{
  short=IM-OTFS,
  long=\ac{im} \ac{otfs},
}
\DeclareAcronym{cdma-otfs}{
  short=CDMA/OTFS,
  long=Code Division Multiple Access \ac{otfs},
}
\DeclareAcronym{dl-cdma-otfs}{
  short=Dl-CDMA/OTFS,
  long=Delay Code Division Multiple Access \ac{otfs},
}
\DeclareAcronym{dp-cdma-otfs}{
  short=Dp-CDMA/OTFS,
  long=Doppler Code Division Multiple Access \ac{otfs},
}
\DeclareAcronym{dd-cdma-otfs}{
  short=DD-CDMA/OTFS,
  long=Delay Doppler Code Division Multiple Access \ac{otfs},
}
\DeclareAcronym{snr}{
  short=SNR,
  long=Signal to Noise Ratio,
}
\DeclareAcronym{eb/no}{
  short=Eb/No,
  long=Energy per bit over Noise power,
}
\DeclareAcronym{crb}{
  short=CRB,
  long=Cramér-Rao Bound,
}
\DeclareAcronym{crlb}{
  short=CRLB,
  long=Cramér-Rao Lower Bound,
}
\DeclareAcronym{papr}{
  short=PAPR,
  long=Peak to Average Power Ratio,
}
\DeclareAcronym{pslr}{
  short=PSLR,
  long=Peak to Side Lobe Ratio,
}
\DeclareAcronym{mimo}{
  short=MIMO,
  long=Multiple Input Multiple Output,
}
\DeclareAcronym{miso}{
  short=MISO,
  long=Multiple Input Single Output,
}
\DeclareAcronym{cp}{
  short=CP,
  long=Cyclic Prefix,
}
\DeclareAcronym{awgn}{
  short=AWGN,
  long=Additive White Gaussian Noise,
}
\DeclareAcronym{far}{
  short=FAR,
  long=Frequency-Agile Radar,
}
\DeclareAcronym{fda}{
  short=FDA,
  long=Frequency-Diverse Array,
}
\DeclareAcronym{ml}{
  short=ML,
  long=Maximum Likelihood,
}
\DeclareAcronym{iot}{
  short=IoT,
  long=Internet of Things,
}
\DeclareAcronym{fh}{
  short=FH,
  long=Frequency Hopping,
}
\DeclareAcronym{pm}{
  short=PM,
  long=Phase Modulation,
}
\DeclareAcronym{cpm}{
  short=CPM,
  long=Continuous Phase Modulation,
}
\DeclareAcronym{am}{
  short=AM,
  long=Amplitude Modulation,
}
\DeclareAcronym{ask}{
  short=ASK,
  long=Amplitude Shift Keying,
}
\DeclareAcronym{fsk}{
  short=FSK,
  long=Frequency Shift Keying,
}
\DeclareAcronym{rf}{
  short=RF,
  long=Radio Frequency,
}
\DeclareAcronym{td}{
  short=TD,
  long=Time Domain,
}
\DeclareAcronym{fd}{
  short=FD,
  long=Frequency Domain,
}
\DeclareAcronym{ad}{
  short=AD,
  long=Antenna Domain,
}
\DeclareAcronym{dd}{
  short=DD,
  long=Delay-Doppler Domain,
}
\DeclareAcronym{tfd}{
  short=TFD,
  long=Time Frequency Domain,
}
\DeclareAcronym{isi}{
  short=ISI,
  long=Inter-Symbol Interference,
}
\DeclareAcronym{gps}{
  short=GPS,
  long=Global Positioning System,
}
\DeclareAcronym{pep}{
  short=PEP,
  long=Pairwise Error Probability,
}
\DeclareAcronym{bpcu}{
  short=bpcu,
  long=Bits Per Channel Use,
}
\DeclareAcronym{med}{
  short=MED,
  long=Minimum Euclidean Distance,
}
\DeclareAcronym{aed}{
  short=AED,
  long=Average Euclidean Distance,
}
\DeclareAcronym{v-blast}{
  short=V-BLAST,
  long=Vertical Bell Laboratories Layered Space-Time,
}
\DeclareAcronym{sm}{
  short=SM,
  long=Spacial Modulation,
}
\DeclareAcronym{mf}{
  short=MF,
  long=Matched Filter,
}
\DeclareAcronym{2s}{
  short=2S,
  long=Two Step,
}
\DeclareAcronym{cdma}{
  short=CDMA,
  long=Code Division Multiple Access,
}
\DeclareAcronym{scma}{
  short=SCMA,
  long=Sparse Code Multiple Access,
}
\DeclareAcronym{noma}{
  short=NOMA,
  long=Non-Orthogonal Multiple Access,
}
\DeclareAcronym{dsss}{
  short=DSSS,
  long=Direct Sequence Spread Spectrum,
}
\DeclareAcronym{iq}{
  short=IQ,
  long=In-phase and Quadrature-phase,
}

\def\BibTeX{{\rm B\kern-.05em{\sc i\kern-.025em b}\kern-.08em
    T\kern-.1667em\lower.7ex\hbox{E}\kern-.125emX}}
\AtBeginDocument{\definecolor{ojcolor}{cmyk}{0.93,0.59,0.15,0.02}}
\def\OJlogo{\vspace{-12pt}\includegraphics[height=24pt]{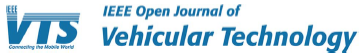}}
\begin{document}
\receiveddate{XX Month, XXXX}
\reviseddate{XX Month, XXXX}
\accepteddate{XX Month, XXXX}
\publisheddate{XX Month, XXXX}
\currentdate{18 October, 2024}
\doiinfo{OJVT.2024.1018000}

\title{\acs*{cdma-otfs} Sensing Outperforms Pure \acs*{otfs} at the Same Communication Throughput}

\author{HUGO HAWKINS\authorrefmark{1}, GRADUATE STUDENT MEMBER, IEEE, CHAO XU\authorrefmark{1}, SENIOR MEMBER, IEEE, LIE-LIANG YANG\authorrefmark{1}, FELLOW, IEEE, AND LAJOS HANZO\authorrefmark{1}, LIFE FELLOW, IEEE}
\affil{School of Electronics and Computer Science, University of Southampton, Southampton SO17 1BJ, UK}
\corresp{CORRESPONDING AUTHOR: Lajos Hanzo (e-mail: lh@ecs.soton.ac.uk).}
\authornote{{L.} Hanzo would like to acknowledge the financial support of the Engineering and Physical Sciences Research Council (EPSRC) projects under grant EP/Y037243/1, EP/W016605/1, EP/X01228X/1, EP/Y026721/1, EP/W032635/1, EP/Y037243/1 and EP/X04047X/1 as well as of the European Research Council's Advanced Fellow Grant QuantCom (Grant No. 789028).}
\markboth{\acs*{cdma-otfs} Sensing Outperforms Pure \acs*{otfs} at the Same Communication Throughput}{Hugo Hawkins \textit{et al.}}

\begin{abstract}
    There is a dearth of publications on the subject of spreading-aided \ac{otfs} solutions, especially for \ac{isac}, even though \ac{cdma} assisted multi-user \ac{otfs} (\acs{cdma-otfs}) exhibits tangible benefits. Hence, this work characterises both the communication \ac{ber} and sensing \ac{rmse} performance of \ac{cdma-otfs}, and contrasts them to pure \ac{otfs}.
    
    Three \ac{cdma-otfs} configurations are considered: \ac{dl-cdma-otfs}, \ac{dp-cdma-otfs}, and \ac{dd-cdma-otfs}, which harness direct sequence spreading along the delay axis, Doppler axis, and \acs{dd} domains respectively. For each configuration, the performance of Gold, Hadamard, and Zadoff-Chu sequences is investigated.
    
    The results demonstrate that Zadoff-Chu \ac{dl-cdma-otfs} and \ac{dd-cdma-otfs} consistently outperform pure \ac{otfs} sensing, whilst maintaining a similar communication performance at the same throughput. The extra modulation complexity of \ac{cdma-otfs} is similar to that of other \ac{otfs} multi-user methodologies, but the demodulation complexity of \ac{cdma-otfs} is lower than that of some other \ac{otfs} multi-user methodologies. \ac{cdma-otfs} sensing can also consistently outperform \ac{otfs} sensing whilst not requiring any additional complexity for target parameter estimation. Therefore, \ac{cdma-otfs} is an appealing candidate for implementing multi-user \ac{otfs} ISAC.
\end{abstract}



\maketitle

\section{INTRODUCTION}

\IEEEPARstart{I}{ngrated} Sensing and Communication (\acs{isac}) is a subject of considerable interest for future wireless generations \cite{JRC_survey_plus_tran_design,MU_MIMO_JCAS_and_ISAC,isac_human_activity_magazine,isac_survey_iot_2024}, as the number of wireless devices is expected to drastically increase. Since the \ac{otfs} concept was first introduced \cite{otfs_first_conf,otfs_second_conf}, its employment for \ac{isac} \cite{otfs_survey-ish_2021} has been a topic of interest. This is due in part to \ac{otfs} being less affected by Doppler shift than \ac{ofdm}, and to the \ac{dd} channel being defined by the delay and Doppler shifts of the propagation paths. This leads to the \ac{dd} channel fluctuating at a slower rate than its \ac{tfd} and \ac{td} counterparts. When the delay and Doppler shifts of the propagation paths are perfectly synchronous with the system's sampling grid, the channel can be modelled using a sparse matrix. This can simplify the associated target parameter estimation algorithms \cite{otfs_radar_basic_conference}.

However, in practice, the delay and Doppler shifts are rarely integer multiples of the system resolutions, hence more complex detection algorithms are required for accurate target estimation. The full-search based \ac{ml} attains the best performance at the highest complexity. Hence, Gaudio et al. \cite{OTFS_JRC_param_est} compare the sensing performance of \ac{ofdm} and \ac{otfs} employing \ac{ml} detection applied to the entire possible set of delay and Doppler indices, at an unspecified resolution. At low velocities, where the Doppler shift is 0.28\% of the subcarrier spacing, \ac{ofdm} \ac{ml} sensing approaches the \ac{crb} at a lower \ac{snr} than \ac{otfs}. This suggests that at low velocities, \ac{ofdm} is more suitable for sensing than \ac{otfs}. However, \cite{OTFS_JRC_param_est} does not compare the two waveforms at high velocities.

In \cite{otfs_rad_frac_inds_diff_est,otfs_com_mp_rad_frac_diff_est}, the authors conceive a two dimensional correlation-based method for \ac{otfs} integer delay and Doppler index estimation. A generalised likelihood ratio test is developed to estimate the number of targets. The fractional indices of the estimated targets are determined by comparing the power-ratio of the indices surrounding the estimated integer indices in the imaging matrix. This method requires each target to have unique integer delay and Doppler indices. The authors observe that the correlation-based method results in an error floor at high \acp{snr} due to the inter-symbol interference caused by the multiple propagation paths in the channel. Tang et al. \cite{otfs_isac_int_ind_mf_frac_ind_fibonacci} propose a two-step sensing methodology to estimate the integer and fractional delay and Doppler indices of the targets for \ac{otfs} sensing. The first step employs a matched filter-based method to determine the integer indices. The second step utilises a Fibonacci-search based algorithm to iteratively determine the fractional indices. Unlike other parameter estimation methods, which reach an error floor at high \acp{snr}, the estimation error of the Fibonacci-search based algorithm continues to decrease as the \ac{snr} increases.

In \cite{otfs_chan_and_rad_param_est_letter}, Muppaneni et al. design a two-step method for sensing the target parameters of \ac{otfs}. Firstly, the integer indices are estimated by maximising a cost function for a fixed number of propagation paths. Secondly, the fractional indices are estimated using a matrix of all the integer indices and the previously estimated fractional indices. For the first path, only integer indices of the other paths are utilised to estimate the first path fractional indices. Once the fractional indices for a path have been estimated, the channel attenuation of the associated path is then estimated. After the fractional indices of a path are estimated, the difference between the received signal and the reconstructed received signal from the estimates is calculated. If this difference becomes smaller than a specific threshold, the sensing procedure is terminated, and path estimation is concluded. Otherwise, the algorithm continues until the fixed number of propagation path variables has been estimated. This method is also harnessed for communication channel estimation using pilot frames. The results presented in \cite{otfs_chan_and_rad_param_est_letter} show that the proposed algorithm leads to a lower sensing estimation error and communication \ac{ber} than the modified \ac{ml} benchmark used for comparison.

A low-complexity variant of \ac{otfs}, known as \ac{otsm}, was first introduced by Thaj et al. \cite{otsm_first}. This waveform modulates the symbols in the delay-sequency (sic) domain, as opposed to the \ac{dd} of \ac{otfs}. This is achieved by replacing the \ac{idft} along the Doppler axis by the Walsh-Hadamard transform. Since this is a purely real-valued transform, a real-valued transmitted signal can be generated by employing a real-valued modulation scheme. The Hadamard transform is also more computationally efficient than the \ac{dft} and the \ac{idft} used for \ac{otfs}. However, since the modulated waveform is constructed in the delay-sequency domain, not the \ac{dd}, the propagation paths with identical delays cannot be separated without more complex detection methods, even if they have different Doppler shifts. Thaj et al. \cite{otsm_first} show that \ac{otsm} has a similar \ac{ber} performance to \ac{otfs}, but it is computationally more efficient due to the use of the Hadamard transform. The pilot power and \ac{papr} of \ac{otsm} is reduced by Neelam and Sahuin \cite{otsm_letter_reduced_papr_and_pilot_power}, by superimposing the pilot symbols on the data, and employing a low complexity iterative channel estimator. Moreover, the \ac{ber} upper bound of \ac{otsm} is determined by Sui et al. \cite{otsm_ber_upper_bound_and_vamp_em_detector}, and a novel vector approximate message passing-based expectation-maximization detector is developed to improve the \ac{ber} performance of \ac{otsm}.

Doosti-Aref et al. \cite{otsm_sequency_im} harness \ac{im} in the sequency domain, to improve both the spectral efficiency and \ac{ber} of \ac{otsm}. This method is further developed in \cite{otsm_parwise_sequency_im}, where pairs of sequence indices of the sequency domain are determined by \ac{im}, as opposed to detecting individual indices. This improves the energy efficiency of the system, and reduces the error propagation imposed by erroneous activated index estimation.

The estimation and compensation of the \ac{iq} imbalance in the received signal caused by the hardware at high frequencies is investigated for \ac{otsm} and \ac{otfs} by Neelam and Sahu \cite{otsm_iq_imbalance_estimation_compensation_receiver}. The estimation is achieved by placing two pilot symbols in the delay-sequency domain transmitted signal, and then estimating the imbalance based on these symbols in the delay-time domain. The addition of a second pilot symbol decreases the number of indices available for data, since guard bands are also required between the pilot symbols, in addition to the guard band between the pilot symbols and the data. The \ac{iq} compensation is implemented in the delay-time domain. The compensation of the imbalance is improved as the pilot power is increased. This work is then also extended to determine both the carrier frequency offset and the channel parameters by Neelam and Sahu \cite{otsm_iq_imbalance_estimation_compensation_receiver_carrier_frequency_offset_channel_estimation}, where both integer delay and fractional Doppler indices are considered.

The work in \cite{otsm_iq_imbalance_estimation_compensation_receiver} is further extended to transmitter and receiver imbalance estimation and compensation as well as channel estimation by the same authors \cite{otsm_iq_imbalance_estimation_compensation_transmitter_receiver_and_channel_estimation_training_sequence}. A complex pseudo noise-based training sequence is placed in the indices allocated to the pilot symbols. The imaginary part of the sequence is a cyclic shift of the real part of the sequence. The channel is estimated at discrete intervals, where the initial estimations are further refined by linear interpolation. An iterative algorithm is conceived for estimating and compensating the \ac{iq} imbalance.

Singh et al. \cite{otsm_iq_imbalance_tx_rx_carrier_frequency_offset_dc_offset_estimation_compensation} implement a deep learning-based detector for \ac{otsm} to compensate for hardware impairments. This deep learning-based detector involves a convolutional neural network, which is trained on the transmitted, received, and \ac{mmse} detector outputs. A data augmentation scheme is then developed to further enhance the input data to the neural network. The resultant detector has a slightly higher complexity than an \ac{mmse} detector, whilst reducing the \ac{ber}.

Reddy et al. \cite{otfs_otsm_sc_comparison} design a multi-user uplink method based on upsampled and circularly shifted \ac{ofdm}. These signals, when combined at the base station, are equivalent to single-user \ac{otfs}, \ac{otsm}, or block-based single carrier signals. This is equivalent to a system that allocates delay indices to users, where the secondary domain of the data can be selected simply by interchanging the matrix multiplying the modulated data. Specifically, the system may opt for the \ac{idft} for \ac{otfs}/Zak-\ac{otfs}, the Hadamard transform for \ac{otsm}, or the identity matrix for single carrier \ac{td} schemes. The \ac{ber} performance of the \ac{otsm} method is comparable to that of the \ac{otfs} method.

Another method applying a real-valued transform to \ac{otfs} is proposed by Kalpage et al. \cite{discrete_cosine_transform_otfs_papr_reduction}. A modified Zak-\ac{otfs} method relying on the \ac{dct} is developed to reduce the \ac{papr} of \ac{otfs}. Replacing the \ac{dft} by the \ac{dct} reduces both the \ac{papr} and the complexity of the system, without eroding the \ac{ber} for the conditions considered by the authors.

Surabhi et al. \cite{otfs_multi_user_DD_mat_split} design a multi-user \ac{otfs} system that splits the available \ac{dd} indices into multiple sub-groups, each of which is allocated to a user. The sub-groups can be along the delay axis, the Doppler axis, or both axes. The \ac{ber} of the schemes relying on index groups along the delay or Doppler axis is lower than that of the schemes that allocate indices for a user along two axes. The \ac{ber} of this system using \ac{ml} or message passing-based detection is lower than that of \ac{ofdma} and \ac{sc-fdma}.

Khammammetti and Mohammed \cite{otfs_multi_user_DD_mat_split_guard_band_between_user_blocks} propose a \ac{dd} index allocation method for multi-user communications similar to that of \cite{otfs_multi_user_DD_mat_split}, but place guard bands between the indices assigned to users, to minimise the inter-user interference. This leads to a simpler detector, at the cost of an eroded throughput.

Ge et al. \cite{otfs_dd_matrix_partitioning_uplink_stationary_and_moving_users} apply the \ac{dd} matrix partitioning method to uplink communication, separating stationary and mobile devices. The stationary devices modulate along the delay axis, and the mobile devices modulate along the Doppler axis. As the base station is static, the stationary users do not experience Doppler shift. A successive interference cancellation aided iterative turbo receiver utilising soft inputs and outputs is developed for detecting the uplink signals. The stationary user signals are decoded, followed by the information arriving from the mobile devices.

\begin{table*}[tp]
    \centering
    {
    \caption{Contrasting contributions to the literature}
    \begin{tabularx}{\linewidth}{|>{\raggedright\arraybackslash}X|c|c|c|c|c|c|c|c|c|c|c|}
        \hline
        \backslashbox{Topics}{Papers} & \cite{otfs_tsma} & \cite{convolutional_scma_otfs_chan_est} & \cite{otfs_scma_uplink_downlink} & \cite{otfs_scma_downlink} & \cite{otfs_otsm_sc_comparison} & \cite{otfs_multi_user_DD_mat_split} & \cite{otfs_ds_delay_spread_comm_only_conf} & \cite{otfs_ds_doppler_spread_comm_only_conf} & \cite{otfs_noma_scma_downlink_conference} & This work\\
        \hline
        \multicolumn{11}{|c|}{Channel modeling}\\
        \hline
        Fractional delay indices & & & & & & & & & & \checkmark\\
        \hline
        Fractional Doppler indices & & & \checkmark & \checkmark & & \checkmark & & & \checkmark & \checkmark\\
        \hline
        \multicolumn{11}{|c|}{Sequence spreading}\\
        \hline
        Sequence spreading & \checkmark & \checkmark & \checkmark & \checkmark & \checkmark & & \checkmark & \checkmark & \checkmark & \checkmark\\
        \hline
        Dense sequences\textsuperscript{1} & \checkmark & & & & \checkmark & & \checkmark & \checkmark & & \checkmark\\
        \hline
        \multicolumn{11}{|c|}{Multi-user methods}\\
        \hline
        Multi-user communication & \checkmark & \checkmark & \checkmark & \checkmark & \checkmark & \checkmark & & & \checkmark & \checkmark\\
        \hline
        Resource allocation on delay only & \checkmark & \checkmark & \checkmark & & & \checkmark & \checkmark & & & \checkmark\\
        \hline
        Resource allocation on Doppler only & & \checkmark & \checkmark & & & \checkmark & & \checkmark & & \checkmark\\
        \hline
        Resource allocation on \acs*{dd} indices & & & & \checkmark & & \checkmark & & & \checkmark & \checkmark\\
        \hline
        Code multiple access methods & \checkmark & \checkmark & \checkmark & \checkmark & & & & & \checkmark & \checkmark\\
        \hline
        \textbf{Code multiple access along the delay only, Doppler only, and \acs*{dd} indices} & & & & & & & & & & \checkmark\\
        \hline
        \multicolumn{11}{|c|}{\acs*{isac}}\\
        \hline
        \textbf{Code multiple access \acs*{otfs} communication and sensing} & & & & & & & & & & \checkmark\\
        \hline
        \textbf{Effect of spreading on the sensing \acs*{rmse} performance} & & & & & & & & & & \checkmark\\
        \hline
        \multicolumn{11}{c}{\textsuperscript{1} Dense sequences: all or the majority of the elements have a non-zero value, as opposed to sparse sequences}\\
    \end{tabularx}
    \label{tab:novelties}
    }
\end{table*}

\ac{scma} has been amalgamated in conjunction with \ac{otfs} in \cite{convolutional_scma_otfs_chan_est,otfs_noma_scma_downlink_conference,otfs_scma_downlink,otfs_scma_uplink_downlink}, relying on \ac{noma} for supporting a higher number of users than the number of available resources blocks, albeit at the cost of increased inter-user interference.

Thomas et al. \cite{convolutional_scma_otfs_chan_est} allocate the modulated symbols along either the delay or Doppler axes using sparse codes. Both the uplink and downlink are considered. In the downlink, a single pilot symbol associated with an appropriate guard band is used for channel estimation, to achieve a communication performance close to the perfect channel estimation case. For the uplink, a sophisticated channel estimation method is developed without excessive guard band overhead, as the effect of the sparse codes on the symbols cannot be separated from the multipath channel effects. Convolutional sparse coding techniques are utilised for uplink channel estimation, as the pilot symbols also rely on sparse codes in the pilot band, surrounded by guard bands.

Wen at al. \cite{otfs_noma_scma_downlink_conference,otfs_scma_downlink} allocate the symbols based on each user's sparse code along the delay and Doppler indices. The system firstly estimates the vector of superimposed transmitted symbols from all users in the \ac{td}, then decodes the symbols gleaned from each user in the \ac{dd} employing a message passing algorithm. The system iterates between the two domains to accurately estimate the transmitted symbols.

Deka et al. \cite{otfs_scma_uplink_downlink} harness \ac{scma} \ac{otfs} for both uplink and downlink communication, with the sparse codes aligned either along the delay axis, or the Doppler axis. Channel estimation is performed by embedding a pilot symbol surrounded by guard bands in the transmitted signal. The results show a lower \ac{ber} for \ac{scma} \ac{otfs} than for power-domain \ac{otfs} \ac{noma} and \ac{dd} index allocation \ac{otfs} for the same normalised user load.

Although \ac{scma}-\ac{otfs} is a promising multi-user method, the characteristics of the transmitted signal are not optimal for sensing, since signals exhibiting a higher degree of randomness lead to an eroded sensing performance \cite{on_fundamental_limits_isac}. The sparsity of the codes is beneficial for reducing the detection and demodulation complexity, but increases the signal variablility. This issue is also observed when IM is harnessed \cite{im-ofdm_isac_outperform_ofdm_isac_collection}.

There are also a few publications on dense sequence spreading \ac{otfs} \cite{otfs_ds_delay_spread_comm_only_conf,otfs_ds_doppler_spread_comm_only_conf,otfs_tsma}. In \cite{otfs_ds_delay_spread_comm_only_conf}, Sun et al. spread the symbols along the delay axis of the \ac{dd} matrix. Two spreading sequence methods are considered: a single spreading sequence used to spread each symbol, or separate sequences used to spread each symbol. In the first case, a sequence having good autocorrelation properties is designed. In the second case, the optimisation problem is too complex, due to the immense variety of possible sequence combinations. Gold and $m$-sequences are used in this case. The \ac{ber} of spread \ac{otfs} is shown to be lower than that of \ac{dsss} or spread \ac{ofdm}.

In Cao et al. \cite{otfs_ds_doppler_spread_comm_only_conf}, each symbol is spread along the Doppler axis using $m$-sequences. A rake receiver is designed to detect the symbols. The channel is assumed to have two paths, each with identical delays. This novel receiver leads to a lower \ac{ber} than conventional \ac{mmse} detection in the limited channel conditions considered. The performance difference is more pronounced if the gains of each path are of similar magnitude.

In Ma et al. \cite{otfs_tsma}, the users are assigned to groups. Each user group is allocated a Doppler index, with each user symbol spread over the delay indices at the user group Doppler index. The sequences utilised are cyclically orthogonal, such as the columns or rows of a \ac{dft} matrix. The spread symbols are interleaved along the delay axis. However, no performance comparison to OTFS is provided.

\subsection{Contributions}
\label{sec:introduction_contributions}

Again, there is a paucity of publications on the subject of dense sequence spreading \ac{otfs}, especially for \ac{isac}. Hence, this work analyses the communication \ac{ber} and sensing \ac{rmse} performance of \ac{cdma-otfs}, and contrasts them with pure \ac{otfs}. Table \ref{tab:novelties} explicitly juxtaposes the novelties of the proposed system to the existing literature, which are detailed below:
\begin{itemize}
    \item A detailed analysis of \ac{cdma-otfs} in the context of \ac{isac} is provided, where both fractional delay indices and fractional Doppler indices are considered.
    \item An in-depth analysis of the communication \ac{ber} and sensing \ac{rmse} performance of delay only, Doppler only, and delay-Doppler sequence spreading for \ac{cdma-otfs}.
    \item This work demonstrates that Zadoff-Chu \ac{dl-cdma-otfs} and \ac{dd-cdma-otfs} are the configurations that consistently outperform pure \ac{otfs} sensing, whilst maintaining a similar communication performance at the same throughput.
\end{itemize}

\begin{table}[tp]
    \centering
    {
    \caption{Notations}
    \begin{tabular}{|l|c|c|}
        \hline
        \multicolumn{1}{|c|}{Definition} & \multicolumn{1}{c|}{Example} & \multicolumn{1}{c|}{Description}\\
        \hhline{|=|=|=|}
        Scalar value & $a$/$A$ & Italics\\
        \hline
        Vector & $\boldsymbol{a}$ & Bold lower case\\
        \hline
        Matrix & $\boldsymbol{A}$ & Bold upper case\\
        \hline
        Vector or matrix transpose & $(\cdot)^{T}$ & \\
        \hline
        Complex conjugate operation & $(\cdot)^{*}$ & \\
        \hline
        Complex conjugate transpose & $(\cdot)^{H}$ & \\
        \hline
        Inverse of a matrix & $(\cdot)^{-1}$ & \\
        \hline
    \end{tabular}
    \label{tab:notations}
    }
\end{table}

The notations used are shown in Table \ref{tab:notations}.

\section{Transmit Signal Model}
\label{sec:tran_sig_mod}

Sequence spreading across the delay, Doppler, or delay and Doppler indices is applied to \ac{otfs}, relying on Gold, Hadamard, and Zadoff-Chu sequences. A sequence of length $M$ is used for delay-domain spreading, of length $N$ for Doppler-domain spreading, and of length $M N$ for \ac{dd} spreading, where $M$ is the number of subcarriers, and $N$ is the number of symbols slots.

Each sequence $\boldsymbol{c}$ is power-normalised, hence:
\begin{equation}
    \boldsymbol{c}^{H} \boldsymbol{c} = 1 \textnormal{ ,} \label{eq:normalised_sequence}
\end{equation}
where $( \cdot )^{H}$ is the Hermitian/complex transpose.

The matrix $\boldsymbol{C}$ containing all of the spreading sequences utilised is:
\begin{equation}
    \boldsymbol{C} = \left(\boldsymbol{c}_{0} \textnormal{, } \boldsymbol{c}_{1} \textnormal{, ..., } \boldsymbol{c}_{N_{mult} - 1} \right) \textnormal{ ,} \label{eq:c_mat}
\end{equation}
where $N_{mult}$ is the number of multiplexed modulated sequences.

The \ac{dd} transmitted signal vector $\boldsymbol{\tilde{x}} \in \mathbb{C}^{M N \times 1}$ is expressed as:
\begin{equation}
    \boldsymbol{\tilde{x}} = \mathfrak{C} \boldsymbol{s} \textnormal{ ,} \label{eq:x_DD_vect}
\end{equation}
where $\boldsymbol{s} \in \mathbb{C}^{N_{s} \times 1}$ is the vector of the modulated symbols, $\mathfrak{C} \in \mathbb{C}^{M N \times N_{s}}$ is the spreading matrix containing the spreading sequences, where the arrangement of the sequences is dependent on the specific \ac{cdma-otfs} scheme implemented. Furthermore, $N_{s}$ denotes the number of symbols transmitted in a frame. The system throughput is $\frac{\beta N_{s}}{M N}$ \ac{bpcu}, where $\beta$ is the number of bits per symbol. The \ac{cp} is assumed to be sufficiently long, and it is perfectly removed from the received signal at the receiver.

\subsection{\acl*{dl-cdma-otfs}}
\label{sec:ma-delay-otfs_comms}

For \ac{dl-cdma-otfs}, each modulated symbol $s_{n_{mult} \textnormal{, } n}$ is multiplied by a spreading sequence $\boldsymbol{c}_{n_{mult}} \in \mathbb{C}^{M \times 1}$, where $n_{mult} = (0$, $1$, ..., $N_{mult} - 1)$, and $n = (0$, $1$, ..., $N - 1)$. The total number of transmitted symbols $N_{s} = N_{mult} N$, and the maximum number of multiplexed sequences is $M$, hence $1 \leq N_{mult} \leq M$.

The \ac{dd} transmitted signal matrix representation $\boldsymbol{\tilde{X}} \in \mathbb{C}^{M \times N}$ is:
\begin{equation}
    \boldsymbol{\tilde{X}} = \boldsymbol{C} \boldsymbol{S} \textnormal{ ,} \label{eq:x_DD_mat_ma_del}
\end{equation}
where $\boldsymbol{C} \in \mathbb{C}^{M \times N_{mult}}$, $\boldsymbol{S} \in \mathbb{C}^{N_{mult} \times N}$, and $\boldsymbol{S}[n_{mult} \textnormal{, } n] = s_{n_{mult} \textnormal{, } n}$.

The columns of $\boldsymbol{\tilde{X}}$ are stacked to create the \ac{dd} transmitted signal vector $\boldsymbol{\tilde{x}}$:
\begin{equation}
    \boldsymbol{\tilde{x}}[n M:(n + 1) M - 1] = \boldsymbol{\tilde{X}}[0:M - 1 \textnormal{, } n] \textnormal{ ,} \label{eq:x_DD_mat_to_vect}
\end{equation}
where $n = (0$, $1$, ..., $N - 1)$.

The symbols can also be directly spread to form $\boldsymbol{\tilde{x}}$, as in \eqref{eq:x_DD_vect}, where:
\begin{equation}
    \boldsymbol{s}[n N_{mult} + n_{mult}] = s_{n_{mult} \textnormal{, } n} \textnormal{ ,} \label{eq:s_vect_ma_del}
\end{equation}
\begin{equation}
    \mathfrak{C} = diag(\boldsymbol{C}) \textnormal{ ,} \label{eq:c_mat_ma_del}
\end{equation}
and $diag( \boldsymbol{C} )$ is the diagonal operator, which creates a matrix whose diagonal elements are $\boldsymbol{C}$.

As an example, for $M = 4$, $N = 3$, $N_{mult} = 2$, $\boldsymbol{C} \in \mathbb{C}^{M \times N_{mult}} = \mathbb{C}^{4 \times 2}$, and $\mathfrak{C}$ is:
\begin{equation}
    \mathfrak{C} = 
    \begin{bmatrix}
        \boldsymbol{C} & \boldsymbol{0}_{4 \times 2} & \boldsymbol{0}_{4 \times 2}\\
        \boldsymbol{0}_{4 \times 2} & \boldsymbol{C} & \boldsymbol{0}_{4 \times 2}\\
        \boldsymbol{0}_{4 \times 2} & \boldsymbol{0}_{4 \times 2} & \boldsymbol{C}
    \end{bmatrix} \textnormal{ ,} \label{eq:c_mat_ma_del_ex}
\end{equation}
where $\boldsymbol{0}_{4 \times 2}$ is a $4 \times 2$ matrix of $0$.

\subsection{\acl*{dp-cdma-otfs}}
\label{sec:ma-doppler-otfs_comms}

For \ac{dp-cdma-otfs}, each modulated symbol $s_{m \textnormal{, } n_{mult}}$ is multiplied by a spreading sequence $\boldsymbol{c}_{n_{mult}} \in \mathbb{C}^{N \times 1}$, where $m = (0$, $1$, ..., $M - 1)$. The total number of transmitted symbols is $N_{s} = N_{mult} M$, and the maximum number of multiplexed sequences is $N$, hence $1 \leq N_{mult} \leq N$.

The \ac{dd} transmitted signal matrix representation $\boldsymbol{\tilde{X}}$ is:
\begin{equation}
    \boldsymbol{\tilde{X}} = \boldsymbol{S} \boldsymbol{C}^{T} \textnormal{ ,} \label{eq:x_DD_mat_ma_dop}
\end{equation}
where $( \cdot )^{T}$ is the transpose operation, $\boldsymbol{C} \in \mathbb{C}^{N \times N_{mult}}$, $\boldsymbol{S} \in \mathbb{C}^{M \times N_{mult}}$, $\boldsymbol{S}[m \textnormal{, } n_{mult}] = s_{m \textnormal{, } n_{mult}}$, and $m = (0$, $1$, ..., $M - 1)$.

The columns of $\boldsymbol{\tilde{X}}$ are stacked to create the \ac{dd} transmitted signal vector $\boldsymbol{\tilde{x}}$, as shown in \eqref{eq:x_DD_mat_to_vect}. The symbols can also be directly spread to form $\boldsymbol{\tilde{x}}$, as in \eqref{eq:x_DD_vect}, where:
\begin{equation}
    \boldsymbol{s}[m N_{mult} + n_{mult}] = s_{m \textnormal{, } n_{mult}} \textnormal{ ,} \label{eq:s_vect_ma_dop}
\end{equation}
\begin{align}
    \mathfrak{C}[n M + m \textnormal{, } m N_{mult}: (m + 1) N_{mult} - 1] = \qquad \nonumber \\
    \boldsymbol{C}[n \textnormal{, } 0:N_{mult} - 1] \textnormal{ ,} \label{eq:c_mat_ma_dop}
\end{align}
with $m = (0$, $1$, ..., $M - 1)$, and $n = (0$, $1$, ..., $N - 1)$.

As an example, for $M = 4$, $N = 3$, $N_{mult} = 2$, $\boldsymbol{C} \in \mathbb{C}^{N \times N_{mult}} = \mathbb{C}^{3 \times 2}$, and $\mathfrak{C}$ is as shown in \eqref{eq:c_mat_ma_dop_ex}.
\begin{figure*}[!b]
\noindent\makebox[\linewidth]{\rule{0.82\paperwidth}{0.4pt}}
\begin{equation}
    \mathfrak{C} = 
    \begin{bmatrix}
        \boldsymbol{C}[0 \textnormal{, } 0] & \boldsymbol{C}[0 \textnormal{, } 1] & 0 & 0 & 0 & 0 & 0 & 0\\
        0 & 0 & \boldsymbol{C}[0 \textnormal{, } 0] & \boldsymbol{C}[0 \textnormal{, } 1] & 0 & 0 & 0 & 0\\
        0 & 0 & 0 & 0 & \boldsymbol{C}[0 \textnormal{, } 0] & \boldsymbol{C}[0 \textnormal{, } 1] & 0 & 0\\
        0 & 0 & 0 & 0 & 0 & 0 & \boldsymbol{C}[0 \textnormal{, } 0] & \boldsymbol{C}[0 \textnormal{, } 1]\\
        \boldsymbol{C}[1 \textnormal{, } 0] & \boldsymbol{C}[1 \textnormal{, } 1] & 0 & 0 & 0 & 0 & 0 & 0\\
        0 & 0 & \boldsymbol{C}[1 \textnormal{, } 0] & \boldsymbol{C}[1 \textnormal{, } 1] & 0 & 0 & 0 & 0\\
        0 & 0 & 0 & 0 & \boldsymbol{C}[1 \textnormal{, } 0] & \boldsymbol{C}[1 \textnormal{, } 1] & 0 & 0\\
        0 & 0 & 0 & 0 & 0 & 0 & \boldsymbol{C}[1 \textnormal{, } 0] & \boldsymbol{C}[1 \textnormal{, } 1]\\
        \boldsymbol{C}[2 \textnormal{, } 0] & \boldsymbol{C}[2 \textnormal{, } 1] & 0 & 0 & 0 & 0 & 0 & 0\\
        0 & 0 & \boldsymbol{C}[2 \textnormal{, } 0] & \boldsymbol{C}[2 \textnormal{, } 1] & 0 & 0 & 0 & 0\\
        0 & 0 & 0 & 0 & \boldsymbol{C}[2 \textnormal{, } 0] & \boldsymbol{C}[2 \textnormal{, } 1] & 0 & 0\\
        0 & 0 & 0 & 0 & 0 & 0 & \boldsymbol{C}[2 \textnormal{, } 0] & \boldsymbol{C}[2 \textnormal{, } 1]\\
    \end{bmatrix} \textnormal{ .} \label{eq:c_mat_ma_dop_ex}
\end{equation}
\end{figure*}

\subsection{\acl*{dd-cdma-otfs}}
\label{sec:ma-dd-otfs_comms}

For \ac{dd-cdma-otfs}, each modulated symbol $s_{n_{mult}}$ is multiplied by a spreading sequence $\boldsymbol{c}_{n_{mult}} \in \mathbb{C}^{M N \times 1}$. The total number of transmitted symbols $N_{s} = N_{mult}$, and the maximum number of multiplexed sequences is $M N$, hence $1 \leq N_{mult} \leq M N$. The \ac{dd-cdma-otfs} scheme can therefore support more users by relying on a greater number of unique spreading sequences than \ac{dl-cdma-otfs} and \ac{dp-cdma-otfs}.

The symbols are directly spread to form $\boldsymbol{\tilde{x}}$, as in \eqref{eq:x_DD_vect}:
\begin{equation}
    \boldsymbol{\tilde{x}} = \sum^{N_{mult} - 1}_{n_{mult} = 0} \boldsymbol{c}_{n_{mult}} s_{n_{mult}} = \boldsymbol{C} \boldsymbol{s} \textnormal{ ,} \label{eq:x_DD_vect_ma_dd}
\end{equation}
where $\boldsymbol{s} \in \mathbb{C}^{N_{mult} \time 1}$, $\boldsymbol{C} \in \mathbb{C}^{M N \times N_{mult}}$, and $\mathfrak{C} = \boldsymbol{C}$.

\section{Channel Model}
\label{sec:chan}

In this section, the channel models are introduced. The generalised \ac{td}, \ac{tfd}, and \ac{dd} channel models are described in Section \ref{sec:chan_gen}. The communication channel parameters are presented in Section \ref{sec:chan_comm}, while the sensing channel parameters are discussed in Section \ref{sec:chan_sen}.

\subsection{Generalised Channel Model}
\label{sec:chan_gen}

It is assumed that there is no external interference during transmission. The transmitted signal is passed through a time-varying and frequency-selective fading channel, as modelled in \cite{otfs_ris_sagin_double_select_chan}. The \ac{dd} representation of the fading channel is:
\begin{equation}
   \tilde{h}(\tau \textnormal{, } \nu) = \sum^{P - 1}_{p = 0} \tilde{h}_{p} \delta(\tau - \tau_{p}) \delta(\nu - \nu_{p}) \textnormal{ ,} \label{eq:h_dd_def_gen}
\end{equation}
where $\tau$ is the delay, $\nu$ is the Doppler shift, $p = [0$, $1$, ..., $P - 1]$ is the propagation path index, and $P$ is the total number of propagation paths. Furthermore, $\tau_{p}$ is the delay associated with the $p$\textsuperscript{th} path, $\nu_{p}$ is the Doppler shift associated with the $p$\textsuperscript{th} path, $\tilde{h}_{p}$ is the fading gain and path loss associated with the $p$\textsuperscript{th} path, and $\delta(\cdot)$ is the Dirac delta function.

When sampled in the \ac{dd}, the channel can be represented by the time-invariant parameter $\tilde{h}_{p}$, the delay index $\tau^{i}$, and the Doppler index $\nu^{i}$. The delay and Doppler indices are defined as:
\begin{align}
   \tau^{i} &= (\Delta f M) \tau \textnormal{ ,} \label{eq:delay_ind_def_gen} \\
   \nu^{i} &= \frac{N}{\Delta f} \nu \textnormal{ ,} \label{eq:doppler_ind_def_gen}
\end{align}
where $\Delta f$ is the subcarrier spacing.

The \ac{td} representation of the fading channel is:
\begin{equation}
   h_{m \textnormal{, } n \textnormal{, } p} = \tilde{h}_{p} e^{j 2 \pi \nu^{i}_{p} \frac{n M + m - \tau^{i}_{p}}{M N}} \textnormal{ ,} \label{eq:h_td_def_gen}
\end{equation}
where $j = \sqrt{-1}$, $\tau^{i}_{p}$ is the delay index associated with the $p$\textsuperscript{th} propagation path, and $\nu^{i}_{p}$ is the Doppler index associated with the $p$\textsuperscript{th} propagation path.

The \ac{td} channel matrix $\boldsymbol{H}_{n} \in \mathbb{C}^{M \times M}$ is:
\begin{equation}
   \boldsymbol{H}_{n}[m \textnormal{, } \lfloor m - \tau^{i}_{p} \rfloor_{M}] = \sum^{P - 1}_{p = 0} h_{m \textnormal{, } n \textnormal{, } p} \textnormal{ ,} \label{eq:h_td}
\end{equation}
where $\lfloor \cdot \rfloor_{M}$ is the modulo $M$ operator.

When the delay indices are assumed to be integers, the \ac{td} received signal $\boldsymbol{y} \in \mathbb{C}^{M \times 1}$ is
\begin{equation}
   \boldsymbol{y}_{n} = \boldsymbol{H}_{n} \boldsymbol{x}_{n} + \boldsymbol{z} \textnormal{ ,} \label{eq:y_td_int_ind}
\end{equation}
where $\boldsymbol{x}_{n} \in \mathbb{C}^{M \times 1}$ is the \ac{td} transmitted signal, and $\boldsymbol{z}$ is the complex \ac{td} \ac{awgn}, with mean $\mu_{z} = 0$ and variance $\sigma_{z}^{2}$, expressed as $\mathcal{N}(\mu_{z} \textnormal{, } \sigma_{z}^{2})$.

When the integer delay index assumption is discarded, $\left( m - \tau^{i}_{p} \right)$ is not an integer for $m = (0$, $1$, ..., $M - 1)$. As matrices do not have fractional indices, the channel must to be modelled differently. A portion of the fading channel is modelled in the \ac{tfd} and then converted to the \ac{td} as follows:
\begin{align}
    \boldsymbol{y}_{n}[m] = \frac{1}{\sqrt{M}} \sum^{P - 1}_{p = 0} h_{m \textnormal{, } n \textnormal{, } p} \sum^{M - 1}_{\bar{m} = 0} \boldsymbol{\bar{x}}_{n}[\bar{m}] e^{j 2 \pi \frac{\left( m - \tau^{i}_{p} \right) \bar{m}}{M}} \nonumber \\
    + \boldsymbol{z}_{n}[m] \textnormal{ ,} \label{eq:y_td_frac_ind}
\end{align}
where $\boldsymbol{\bar{x}} \in \mathbb{C}^{M \times N}$ is the \ac{tfd} transmitted signal, and $\bar{m} = [0$, $1$, ..., $M - 1]$.

The \ac{tfd} channel matrix  $\boldsymbol{\bar{H}}_{n} \in \mathbb{C}^{M \times M}$ is:
\begin{equation}
   \boldsymbol{\bar{H}}_{n} = \mathcal{F}_{M} \boldsymbol{H}_{n} \mathcal{F}_{M}^{H} \textnormal{ ,} \label{eq:h_tf}
\end{equation}
where $\mathcal{F}_{M}$ is the $M$-point \ac{dft} and $\mathcal{F}_{M}^{H}$ is the $M$-point \ac{idft}.

The \ac{dd} channel matrix $\boldsymbol{\tilde{H}} \in \mathbb{C}^{M N \times M N}$ is:
\begin{equation}
   \boldsymbol{\tilde{H}} = \left( \mathcal{F}_{N} \otimes \mathcal{F}_{M}^{H} \right) \boldsymbol{\bar{H}}_{X} \left( \mathcal{F}_{N}^{H} \otimes \mathcal{F}_{M} \right) \textnormal{ ,} \label{eq:h_dd}
\end{equation}
where $\otimes$ is the Kronecker product, and
\begin{equation}
   \boldsymbol{\bar{H}}_{X} = diag \left( \boldsymbol{\bar{H}}_{n} \right) =
   \begin{bmatrix}
        \boldsymbol{\bar{H}}_{0} & \boldsymbol{0}_{M \times M} & \cdots & \boldsymbol{0}_{M \times M}\\
        \boldsymbol{0}_{M \times M} & \boldsymbol{\bar{H}}_{1} & \cdots & \boldsymbol{0}_{M \times M}\\
        \vdots & \vdots & \ddots & \vdots\\
        \boldsymbol{0}_{M \times M} & \boldsymbol{0}_{M \times M} & \cdots & \boldsymbol{\bar{H}}_{N - 1}
    \end{bmatrix} \textnormal{ .} \label{eq:h_tf_expanded}
\end{equation}

\subsection{Communication Channel Parameters}
\label{sec:chan_comm}

Integer delay indices, fractional Doppler indices, and Rician fading are assumed for communication. There are $P_{com}$ propagation paths. The first path $p_{com} = 0$ is the \ac{los} path, and the remaining $P_{com} - 1$ paths are \ac{nlos} paths. The fading gain $\tilde{h}_{p \textnormal{, } com}$ of the $p_{com}$\textsuperscript{th} propagation path is:
\begin{equation}
    \tilde{h}_{p \textnormal{, } com} =
\begin{cases}
    \sqrt{\frac{\kappa_{com}}{\kappa_{com} + 1}} \textnormal{, } & \textnormal{if } p_{com} = 0\\
    \sqrt{\frac{1}{(\kappa_{com} + 1) (P_{com} - 1)}} \zeta_{p \textnormal{, } com} \textnormal{, } & \textnormal{if } p_{com} > 0 \textnormal{ ,}
\end{cases}
\label{eq:path_loss_com}
\end{equation}
where $\kappa_{com}$ is the Rician K factor, and $\zeta_{p \textnormal{, } com}$ is a complex Gaussian random variable with mean $\mu_{com} = 0$ and variance $\sigma_{com}^{2} = 1$, expressed as $\mathcal{N}(\mu_{com} \textnormal{, } \sigma_{com}^{2})$.

The integer delay index $\tau^{i}_{p \textnormal{, } com}$ is:
\begin{equation}
    \tau^{i}_{p \textnormal{, } com} =
\begin{cases}
    0 \textnormal{,} & \textnormal{if } p_{com} = 0\\
    p_{com} \mathbin{\%} L_{com} \textnormal{,} & \textnormal{if } p_{com} > 0 \textnormal{ \& } P_{com} \geq L_{com}\\
    \lfloor L_{com} \eta_{\tau \textnormal{, } com} \rceil \textnormal{,} & \textnormal{if } p_{com} > 0 \textnormal{ \& } P_{com} < L_{com} \textnormal{,}
\end{cases}
\label{eq:delay_ind_com}
\end{equation}
where $L_{com}$ is the number of delay taps, $p_{com} = [0$, $1$, ..., $P_{com} - 1]$, $\mathbin{\%}$ is the remainder or modulus operator, $\eta_{\tau \textnormal{, } com}$ is a random variable following a uniform distribution between 0 and 1, and $\lfloor \cdot \rceil$ is the rounding function.

If $P_{com} < L_{com}$, no pair of propagation paths will have the same delay index, yielding: $\tau^{i}_{p_{1} \textnormal{, } com} \neq \tau^{i}_{p_{2} \textnormal{, } com}$, where $p_{1} = [0$, $1$, ... $P_{com} - 1]$, $p_{2} = [0$, $1$, ... $P_{com} - 1]$, and $p_{1} \neq p_{2}$.

The fractional Doppler index $\nu^{i}_{p \textnormal{, } com}$ is:
\begin{equation}
    \nu^{i}_{p \textnormal{, } com} =
\begin{cases}
    \nu^{i}_{com \textnormal{, } max} \textnormal{, } & \textnormal{if } p_{com} = 0\\
    \lfloor 2 \nu^{i}_{com \textnormal{, } max} \left( \eta_{\nu \textnormal{, } com} - 0.5 \right) \rceil \textnormal{, } & \textnormal{if } p_{com} > 0 \textnormal{ ,}
\end{cases}
\label{eq:doppler_ind_com}
\end{equation}
where $\eta_{\nu \textnormal{, } com}$ is a random variable following a uniform distribution between 0 and 1, and $\nu^{i}_{com \textnormal{, } max}$ is the maximum integer Doppler index:
\begin{equation}
    \nu^{i}_{com \textnormal{, } max} = \lceil \frac{f_{c} N V_{com}}{\Delta f c_{0}} \rceil \textnormal{ ,} \label{eq:doppler_ind_com_max}
\end{equation}
with $V_{com}$ representing the velocity of the communication receiver, $f_{c}$ is the carrier frequency, $c_{0}$ is the speed of light, and $\lceil \cdot \rceil$ is the ceiling function.

\subsection{Sensing Channel Parameters}
\label{sec:chan_sen}

The delay and Doppler indices for sensing, $\tau^{i}_{p \textnormal{, } sen}$ and $\nu^{i}_{p \textnormal{, } sen}$ respectively, are:
\begin{equation}
    \tau^{i}_{p \textnormal{, } sen} = \frac{2 \Delta f M R_{p \textnormal{, } sen}}{c_{0}} \textnormal{ ,} \label{eq:delay_ind_sen}
\end{equation}
\begin{equation}
    \nu^{i}_{p \textnormal{, } sen} = \frac{2 f_{c} N V_{p \textnormal{, } sen}}{\Delta f c_{0}} \textnormal{ ,} \label{eq:doppler_ind_sen}
\end{equation}
where $R_{p \textnormal{, } sen}$ is the range of the $p_{sen}$\textsuperscript{th} path, $V_{p \textnormal{, } sen}$ is the velocity of the $p_{sen}$\textsuperscript{th} path, $p_{sen} = [0$, $1$, ..., $P_{sen} - 1]$, and $P_{sen}$ is the number of sensing propagation paths.

As monostatic sensing is assumed, the transmitted signal is reflected from the target to the sensing receiver attached or adjacent to the transmitter, hence a factor of 2 is present in \eqref{eq:delay_ind_sen} and \eqref{eq:doppler_ind_sen}. The first $P_{t}$ propagation paths are \ac{los} paths associated with each sensing target, and the remaining $P_{n}$ paths are \ac{nlos} paths. The total number of sensing propagation paths $P_{sen}$ is:
\begin{equation}
    P_{sen} = P_{t} + P_{n} \textnormal{ .} \label{eq:p_sen}
\end{equation}

The fading gain $\tilde{h}_{p_{t} \textnormal{, } sen}$ of the $p_{t}$\textsuperscript{th} target is:
\begin{equation}
    \tilde{h}_{p_{t} \textnormal{, } sen} = \sqrt{\frac{\kappa_{sen}}{\kappa_{sen} + 1}} \sqrt{\alpha_{p_{t}}} \textnormal{ ,} \label{eq:path_loss_sen_target}
\end{equation}
where $\kappa_{sen}$ is the sensing Rician K factor, and $\alpha_{p_{t}}$ is the power gain associated with the $p_{t}$\textsuperscript{th} \ac{los} path, defined as:
\begin{equation}
    \alpha_{p_{t}} = \frac{c_{0}^{2} \sigma_{p_{t}}}{(4 \pi)^{3} f_{c}^{2} R_{p \textnormal{, } sen}^{4}} \textnormal{ ,} \label{eq:power_att_sen_target}
\end{equation}
where $\sigma_{p_{t}}$ is the radar cross-section of the $p_{t}$\textsuperscript{th} target.

The fading gain $\tilde{h}_{p_{n} \textnormal{, } sen}$ of the $p_{n}$\textsuperscript{th} \ac{nlos} path is:
\begin{equation}
    \tilde{h}_{p_{n} \textnormal{, } sen} = \sqrt{\frac{1}{P_{n} (\kappa_{sen} + 1)}} \zeta_{p_{n} \textnormal{, } sen} \min_{\forall p_{t}} \left( \sqrt{\alpha_{p_{t}}} \right) \textnormal{ ,} \label{eq:path_loss_sen_nlos}
\end{equation}
where $\zeta_{p_{n} \textnormal{, } sen}$ is a complex Gaussian random variable with mean $\mu_{sen} = 0$ and variance $\sigma_{sen}^{2} = 1$, expressed as $\mathcal{N}(\mu_{sen} \textnormal{, } \sigma_{sen}^{2})$.

The \ac{nlos} power is set relative to the smallest value of $\alpha_{p_{t}}$, which is associated with the weakest target signal. The smallest value of $\alpha_{p_{t}}$ is used to ensure that no \ac{nlos} paths have an average power higher than $\sqrt{\frac{1}{P_{n} (\kappa_{sen} + 1)}}$ relative to any of the targets. As the system is operating in the mmWave band, the \ac{nlos} reflections are assumed, on average, to be weaker than the \ac{los} signals.

The range and velocity of the \ac{nlos} paths, $R_{p_{n} \textnormal{, } sen}$ and $V_{p_{n} \textnormal{, } sen}$, are:
\begin{equation}
    R_{p_{n} \textnormal{, } sen} = R_{n \textnormal{, } max} \eta_{\tau \textnormal{, } sen} \textnormal{ ,} \label{eq:r_nlos_sen}
\end{equation}
\begin{equation}
    V_{p_{n} \textnormal{, } sen} = 2 V_{n \textnormal{, } max} \left( \eta_{\nu \textnormal{, } sen} - 0.5 \right) \textnormal{ ,} \label{eq:v_nlos_sen}
\end{equation}
where $\eta_{\tau \textnormal{, } sen}$ and $\eta_{\nu \textnormal{, } sen}$ are random variables following a uniform distribution between 0 and 1, and:
\begin{equation}
    R_{n \textnormal{, } max} = \sqrt[4]{\kappa_{sen}} \max_{\forall p_{t}} \left( R_{p_{t} \textnormal{, } sen} \right) \textnormal{ ,} \label{eq:r_nlos_sen_max}
\end{equation}
\begin{equation}
    V_{n \textnormal{, } max} = \frac{\Delta f c_{0}}{4 f_{c}} \textnormal{ ,} \label{eq:v_nlos_sen_max}
\end{equation}

The fourth root of $\kappa_{sen}$ is present in \eqref{eq:r_nlos_sen_max} as the power gain is inversely proportional to $R^{4}$, as seen in \eqref{eq:power_att_sen_target}. A maximum range is fixed, because the reflected signals having \ac{nlos} paths associated with delays larger than the maximum in \eqref{eq:r_nlos_sen_max} are assumed to not significantly interfere, due to the high attenuation associated with a greater range.

\section{Received Signal Processing}
\label{sec:rec_sig_proc}

\subsection{Communication Data Detection}
\label{sec:rec_sig_proc_comm_demod}

The \ac{dd} received signal $\boldsymbol{\tilde{y}} \in \mathbb{C}^{M N \times 1}$ can be represented as:
\begin{equation}
    \boldsymbol{\tilde{y}} = \boldsymbol{\tilde{H}} \boldsymbol{\tilde{x}} + \boldsymbol{\tilde{z}} \textnormal{ ,} \label{eq:y_DD}
\end{equation}
where and $\boldsymbol{\tilde{z}}$ is the complex-valued \ac{awgn} in the \ac{dd}.

\ac{mmse} demodulation is applied at the receiver, with perfect channel estimation assumed. The vector of estimated symbols $\boldsymbol{\hat{s}} \in \mathbb{C}^{N_{s} \times 1}$ is formulated as:
\begin{equation}
    \boldsymbol{\hat{s}} = \boldsymbol{\tilde{G}}^{H} \boldsymbol{\tilde{y}} \textnormal{ ,} \label{eq:mmse_est}
\end{equation}
where $\boldsymbol{\tilde{G}}$ is:
\begin{equation}
    \boldsymbol{\tilde{G}} = \left( \boldsymbol{\tilde{H}} \mathfrak{C} \mathfrak{C}^{H} \boldsymbol{\tilde{H}}^{H} + N_{0} \boldsymbol{I}_{M N \times M N} \right)^{-1} \boldsymbol{\tilde{H}} \mathfrak{C} \textnormal{ ,} \label{eq:mmse_mat}
\end{equation}
where $N_{0}$ is the \ac{awgn} power, and $\boldsymbol{I}_{M N \times M N}$ is the $M N \times M N$ identity matrix. The estimated symbols are then demodulated to obtain the estimated bits.

\subsection{Sensing Target Parameter Estimation}
\label{sec:rec_sig_proc_sen_est}

For sensing, it is assumed that the number and directions of the targets are known. A two step sensing method is applied. The first step utilises a data cancellation-based method to estimate the integer indices of the targets. The accuracy of this step is limited by the system parameters. The second step employs \ac{ml} detection to estimate the fractional component of the indices estimated in step 1. The resolution of the fractional index estimation is a separate system parameter $N_{ML}$.

The first step is a modified data cancellation method adapted from \cite{otfs_radar_basic_conference}. The \ac{dd} received signal can be represented as in \eqref{eq:y_DD}, or as:
\begin{equation}
    \boldsymbol{\tilde{y}} = \boldsymbol{\tilde{X}}_{X} \boldsymbol{\tilde{h}} + \boldsymbol{\tilde{z}} \textnormal{ ,} \label{eq:y_DD_big_x}
\end{equation}
where $\boldsymbol{\tilde{h}} \in \mathbb{C}^{M N \times 1}$ is the vector containing the propagation path gains at the associated delay-Doppler indices, and $\boldsymbol{\tilde{X}}_{X} \in \mathbb{C}^{M N \times M N}$ is an expanded matrix of the \ac{dd} transmitted signal matrix $\boldsymbol{\tilde{X}}$ shown in \eqref{eq:sen_large_x_dc},
\begin{figure*}[!b]
\noindent\makebox[\linewidth]{\rule{0.82\paperwidth}{0.4pt}}
\begin{equation}
    \boldsymbol{\tilde{X}}_{X}[\tau^{i}_{1} + M \nu^{i}_{1} \textnormal{, } \tau^{i}_{2} + M \nu^{i}_{2}] =
\begin{cases}
    \boldsymbol{\tilde{X}}[\tau^{i}_{1} - \tau^{i}_{2} + M \textnormal{, } \nu^{i}_{diff}] e^{\frac{j 2 \pi \nu^{i}_{2} \left( \tau^{i}_{1} - \tau^{i}_{2} \right)}{M N}} \textnormal{, } & \textnormal{if } \tau^{i}_{1} - \tau^{i}_{2} < 0\\
    \boldsymbol{\tilde{X}}[\tau^{i}_{1} - \tau^{i}_{2} \textnormal{, } \nu^{i}_{diff}] e^{\frac{j 2 \pi \nu^{i}_{2} \left( \tau^{i}_{1} - \tau^{i}_{2} \right)}{M N}} \textnormal{, } & \textnormal{if } \tau^{i}_{1} - \tau^{i}_{2} > 0 \textnormal{ ,}
\end{cases}
\label{eq:sen_large_x_dc}
\end{equation}
\end{figure*}
where $\tau^{i}_{1}$ and $\tau^{i}_{2} = [0$, $1$, ..., $M - 1]$, $\nu^{i}_{1}$ and $\nu^{i}_{2} = [0$, $1$, ..., $N - 1]$, and:
\begin{equation}
    \nu^{i}_{diff} =
\begin{cases}
    \nu^{i}_{1} - \nu^{i}_{2} + N \textnormal{, } & \textnormal{if } \nu^{i}_{1} - \nu^{i}_{2} < 0\\
    \nu^{i}_{1} - \nu^{i}_{2} \textnormal{, } & \textnormal{if } \nu^{i}_{1} - \nu^{i}_{2} > 0 \textnormal{ .}
\end{cases}
\label{eq:sen_nu_diff}
\end{equation}

The estimated channel parameter vector $\boldsymbol{\hat{h}}$ is calculated as follows:
\begin{equation}
    \boldsymbol{\hat{h}} = \boldsymbol{\tilde{X}}_{X}^{H} \boldsymbol{\tilde{y}} \textnormal{ .} \label{eq:sen_mf_peak_calc}
\end{equation}

The indices at which the $P_{t}$ peak amplitudes of $||\boldsymbol{\hat{h}}||^{2}$ occur are the estimated integer delay and Doppler indices $\hat{\tau}^{dc}_{p_{t}}$ and $\hat{\nu}^{dc}_{p_{t}}$ of the first step, where $|| \cdot ||$ is the Euclidean norm.

The \ac{ml} second step used for fractional index estimation applies \ac{ml} estimation to the indices adjacent to the $P_{t}$ delay and Doppler integer indices gleaned from the first step. The delay indices $\boldsymbol{\tau}^{ml}_{p_{t}}$ and Doppler indices $\boldsymbol{\nu}^{ml}_{p_{t}}$ considered for the $p_{t}$\textsuperscript{th} target are:
\begin{equation}
    \boldsymbol{\tau}^{ml}_{p_{t}} = \hat{\tau}^{dc}_{p_{t}} \boldsymbol{1}_{(2 N_{ML} + 1) \times 1} + \frac{\boldsymbol{n_{\tau}}}{N_{ML}} \textnormal{ ,} \label{eq:ml_del_indices}
\end{equation}
\begin{equation}
    \boldsymbol{\nu}^{ml}_{p_{t}} = \hat{\nu}^{dc}_{p_{t}} \boldsymbol{1}_{(2 N_{ML} + 1) \times 1} + \frac{\boldsymbol{n_{\nu}}}{N_{ML}} \textnormal{ ,} \label{eq:ml_dop_indices}
\end{equation}
where $\boldsymbol{1}_{(2 N_{ML} + 1) \times 1}$ is a $\left[ (2 N_{ML} + 1) \times 1 \right]$ vector of 1, $\boldsymbol{n_{\tau}} = \big[ -N_{ML}$, $-N_{ML} + 1,$ ..., $0$, ..., $N_{ML} - 1$, $N_{ML} \big]$, $\boldsymbol{n_{\nu}} = \big[ -N_{ML}$, $-N_{ML} + 1$, ..., $0$, ..., $N_{ML} - 1$, $N_{ML} \big]$, and $N_{ML}$ is the interpolation or resolution refinement factor.

The \ac{ml} algorithm creates a set of equivalent channels $\boldsymbol{\hat{H}}_{\tau^{ml}_{p_{t}} \textnormal{, } \nu^{ml}_{p_{t}}}$, with a gain of 1, for the sets $\boldsymbol{\tau}^{ml}_{p_{t}}$ and $\boldsymbol{\nu}^{ml}_{p_{t}}$. The gain is set to 1 since no channel gain estimation is performed. The algorithm calculates the peak associated with each combination, and determines the indices $\hat{\tau}^{ml}_{p_{t}}$ and $\hat{\nu}^{ml}_{p_{t}}$ estimated by selecting the combination with the largest peak, following \cite{OTFS_and_OFDM_JRC_performance_analysis}:
\begin{equation}
    (\hat{\tau}^{ml}_{p_{t}} \textnormal{, } \hat{\nu}^{ml}_{p_{t}}) = \arg \max\limits_{\forall \tau^{ml}_{p_{t}} \textnormal{, } \nu^{ml}_{p_{t}}} \frac{||\boldsymbol{\tilde{x}}^{H} \boldsymbol{\hat{H}}_{\tau^{ml}_{p_{t}} \textnormal{, } \nu^{ml}_{p_{t}}}^{H} \boldsymbol{\tilde{y}}||^{2}}{\boldsymbol{\tilde{x}}^{H} \boldsymbol{\hat{H}}_{\tau^{ml}_{p_{t}} \textnormal{, } \nu^{ml}_{p_{t}}}^{H} \boldsymbol{\hat{H}}_{\tau^{ml}_{p_{t}} \textnormal{, } \nu^{ml}_{p_{t}}} \boldsymbol{\tilde{x}}} \textnormal{ ,} \label{eq:sen_ml_peak_calc}
\end{equation}
where $\boldsymbol{\tilde{y}}$ is the \ac{dd} received signal vector.

The indices $\hat{\tau}^{ml}_{p_{t}}$ and $\hat{\nu}^{ml}_{p_{t}}$ are then used to calculate the target range $\hat{R}_{p_{t} \textnormal{, } sen}$ and velocity $\hat{V}_{p_{t} \textnormal{, } sen}$, following \cite{im-ofdm_isac_outperform_ofdm_isac_collection}:
\begin{equation}
    \hat{R}_{p_{t} \textnormal{, } sen} = \frac{\hat{\tau}^{ml}_{p_{t}} c_{0}}{2 M f_{s}} \textnormal{ ,} \label{eq:r_sen_est}
\end{equation}
\begin{equation}
    \hat{V}_{p_{t} \textnormal{, } sen} = \frac{\hat{\nu}^{ml}_{p_{t}} \Delta f c_{0}}{2 N f_{c}} \textnormal{ .} \label{eq:v_sen_est}
\end{equation}

\section{\acl*{crb}}
\label{sec:crb}

Following \cite{OTFS_CRB}, the average unbiased estimator \acp{crb} for the associated range and velocity estimation are defined as:
\begin{align}
    \hat{\sigma}_{l \textnormal{, } R}^{2} &\geq \frac{N_{0}}{P_{avg} |\tilde{h}_{p_{t} \textnormal{, } sen}|^{2} \pi^{2} M N (M - 1)^{2}} \left( \frac{c_{0}}{2 \Delta f} \right)^{2} \textnormal{ ,} \label{eq:crb_ran_var} \\
    \hat{\sigma}_{l \textnormal{, } V}^{2} &\geq \frac{N_{0}}{P_{avg} |\tilde{h}_{p_{t} \textnormal{, } sen}|^{2} \pi^{2} M N (N - 1)^{2}} \left( \frac{c_{0} \Delta f}{2 f_{c}} \right)^{2} \textnormal{ ,} \label{eq:crb_vel_var}
\end{align}
where $\hat{\sigma}_{l \textnormal{, } R}^{2}$ and $\hat{\sigma}_{l \textnormal{, } V}^{2}$ are the variance of the target range and of the velocity estimation errors, respectively, while $P_{avg}$ is the average power of the transmitted signal.

As the \ac{rmse} is the metric used for sensing, the standard deviation is utilised:
\begin{align}
    \hat{\sigma}_{l \textnormal{, } R} &\geq \sqrt{\frac{N_{0}}{P_{avg} |\tilde{h}_{p_{t} \textnormal{, } sen}|^{2} \pi^{2} M N (M - 1)^{2}}} \frac{c_{0}}{2 \Delta f} \label{eq:crb_ran_rmse} \textnormal{ ,} \\
    \hat{\sigma}_{l \textnormal{, } V} &\geq \sqrt{\frac{N_{0}}{P_{avg} |\tilde{h}_{p_{t} \textnormal{, } sen}|^{2} \pi^{2} M N (N - 1)^{2}}} \frac{c_{0} \Delta f}{2 f_{c}} \textnormal{ ,} \label{eq:crb_vel_rmse}
\end{align}
where $\hat{\sigma}_{l \textnormal{, } R}$ and $\hat{\sigma}_{l \textnormal{, } V}$ are the standard deviations of the target range and velocity estimation errors, respectively.

It is important to note that this \ac{crb} is an average \ac{crb}, not a true lower bound.

\section{Computational Complexity Analysis}
\label{sec:compx_anlys}

This section quantifies the additional complexity introduced by \ac{cdma-otfs} to the modulation and communication demodulation compared to \ac{otfs}, and the complexity of the target parameter estimation methods.

\subsection{Additional Modulation and Demodulation Complexity Compared to \acs*{otfs}}
\label{sec:compx_anlys_mod_demod}

In \eqref{eq:x_DD_vect}, the modulation relies on a multiplication of an $(M N \times N_{s})$ matrix by an $(N_{s} \times 1)$ vector. The computational complexity of this operation is on the order of $O(M N N_{s})$, for all the \ac{cdma-otfs} schemes.

For \ac{dl-cdma-otfs}, $N_{s} = N_{mult} N$, $1 \leq N_{mult} \leq M$, hence the complexity at the minimum and maximum throughputs is $O(M N^{2})$ and $O(M^{2} N^{2})$, respectively.

For \ac{dp-cdma-otfs}, $N_{s} = N_{mult} M$, $1 \leq N_{mult} \leq N$, hence the complexity at the minimum and maximum throughputs is $O(M^{2} N)$ and $O(M^{2} N^{2})$, respectively.

For \ac{dd-cdma-otfs}, $N_{s} = N_{mult}$, $1 \leq N_{mult} \leq M N$, hence the complexity at the minimum and maximum throughputs is $O(M N)$ and $O(M^{2} N^{2})$, respectively.

The computational complexity of all the \ac{cdma-otfs} schemes is the same for the same throughput. Modulating using \eqref{eq:x_DD_mat_ma_del} for \ac{dl-cdma-otfs} and \eqref{eq:x_DD_mat_ma_dop} for \ac{dp-cdma-otfs} leads to the same complexity.

For \ac{mmse} demodulation, there is an additional matrix multiplication required in \eqref{eq:mmse_mat}, where the channel matrix $(M N \times M N)$ is multiplied by the spreading matrix $(M N \times N_{s})$. This multiplication can be performed once, with the resulting matrix used three times in \eqref{eq:mmse_mat}. The complexity of the multiplication is on the order of $O(M^{2} N^{2} N_{s})$.

\subsection{Complexity of the Sensing Methods}
\label{sec:compx_anlys_sen}

There are $M N \times M N$ operations required for the creation of the expanded matrix of the \ac{dd} transmitted signal \eqref{eq:sen_large_x_dc}, hence the complexity is $ O(M^{2} N^{2})$. The complexity of the matrix multiplication of the data cancellation method \eqref{eq:sen_mf_peak_calc} is $O(M N \times M N \times 1) = O(M^{2} N^{2})$, since an $(M N \times M N)$ matrix is multiplied by an $(M N \times 1)$ vector. The total complexity of this algorithm is therefore $O(2 M^{2} N^{2})$.

The computational complexity of \eqref{eq:sen_ml_peak_calc} for fractional index estimation is $O(2 M N)$. This operation is repeated $(2 N_{ML} + 1)^{2}$ times, hence the total complexity of the \ac{ml} fractional index estimation is $O(2 M N (2 N_{ML} + 1)^{2})$.

\subsection{Discussions}
\label{sec:compx_anlys_disc}

The additional complexity imposed on the modulation by \ac{cdma-otfs} is comparable to that of \ac{scma} \ac{otfs} schemes, as both involve the spreading of symbols across the available resources. There exist certain matrix multiplication methods having reduced complexity for sparse matrices, so the additional complexity of \ac{scma} \ac{otfs} may be reduced, depending on the spreading matrix structure. Other multi-user methods, such as directly allocating \ac{dd} indices to users \cite{otfs_multi_user_DD_mat_split,otfs_multi_user_DD_mat_split_guard_band_between_user_blocks,otfs_dd_matrix_partitioning_uplink_stationary_and_moving_users}, may have a lower modulation complexity, depending on their modulation structure.

The additional demodulation complexity of \ac{cdma-otfs} is much lower than that of \ac{dd} index allocation, because the properties of the dense sequences chosen for \ac{cdma-otfs} allow for a relatively simple demodulation, due to their desirable cross-correlation properties. Although no correlation operations are harnessed for communication detection and demodulation, the low cross-correlation of the sequences mitigates inter-user/inter-symbol inteference. This trend becomes more pronouced when perfectly orthogonal sequences having zero-cross-correlation are employed. \ac{dd} index allocation requires more complex demodulation methods, for example, iterative methods such as successive interference cancellation, in order to mitigate the inter-symbol and inter-user interference.

\section{Simulation Results and Discussions}
\label{sec:results}

This section discusses the \ac{ber} results and sensing \ac{rmse} results of both multi-user \ac{cdma-otfs} and single-user \ac{otfs}. The single-user \ac{otfs} system assigns a single symbol to each \ac{dd} resource, hence transmitting $M N$ symbols in a frame. In the legend, ``Gold'' refers to Gold sequence spreading, ``Had'' refers to Hadamard sequence spreading, and ``ZC'' refers to Zadoff-Chu sequence spreading.

\subsection{Communication \acs*{ber} Results}
\label{sec:results_ber}

\begin{table}[tp]
    \centering
    {
    \caption{Communication simulation parameters}
    \begin{tabular}{|l|r|}
        \hline
        \multicolumn{1}{|c|}{Variable} & \multicolumn{1}{|c|}{Value}\\
        \hhline{|=|=|}
        Carrier frequency $f_{c}$ & $40$ GHz\\
        \hline
        Subcarrier spacing $\Delta f$ & $120$ kHz\\
        \hline
        Number of communication delay taps $L_{com}$ & $3$\\
        \hline
        Number of communication propagation paths $P_{com}$ & $3$\\
        \hline
        Rician K factor $\kappa_{com}$ & $0$ dB\\
        \hline
        Number of subcarriers $M$ & $64$\\
        \hline
        Number of symbols sent per frame $N$ & $64$\\
        \hline
        Communication receiver velocity & $200$ m/s\\
        \hline
        Minimum number of bit errors & $600$\\
        \hline
        Maximum number of bits simulated & $1 \times 10^{7}$\\
        \hline
    \end{tabular}
    \label{tab:com_sim_vars}
    }
\end{table}

\begin{figure*}[tp]
    \centering
    \begin{subfigure}{.45\textwidth}
        \includegraphics[width=0.9\linewidth]{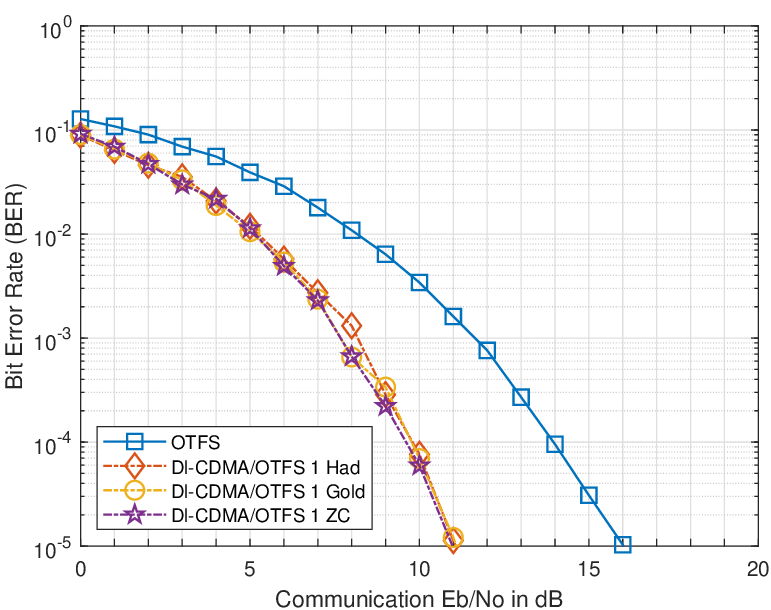}
        \caption{\acs*{otfs} and \acs*{dl-cdma-otfs}}
        \label{fig:ber_n_mult_1_del_only}
    \end{subfigure}
    \begin{subfigure}{.45\textwidth}
        \includegraphics[width=0.9\linewidth]{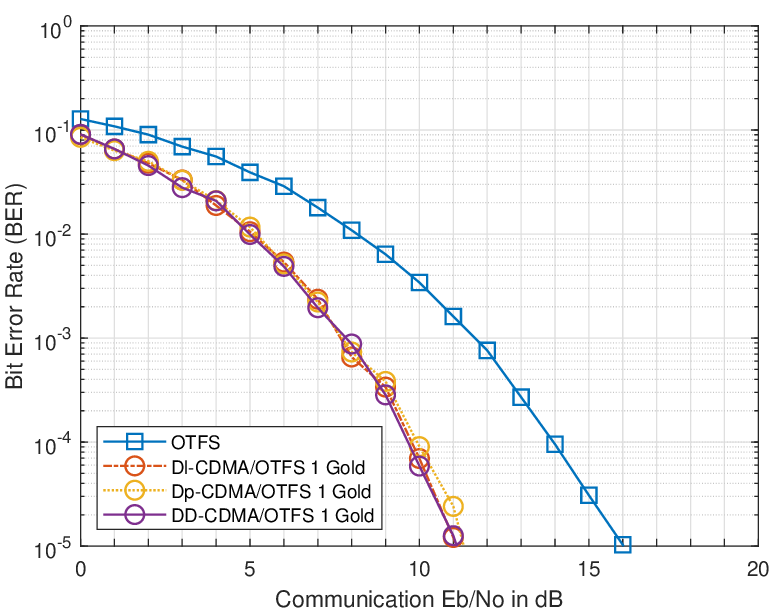}
        \caption{\acs*{otfs} and Gold sequence}
        \label{fig:ber_n_mult_1_gold_only}
    \end{subfigure}
    \begin{subfigure}{.45\textwidth}
        \includegraphics[width=0.9\linewidth]{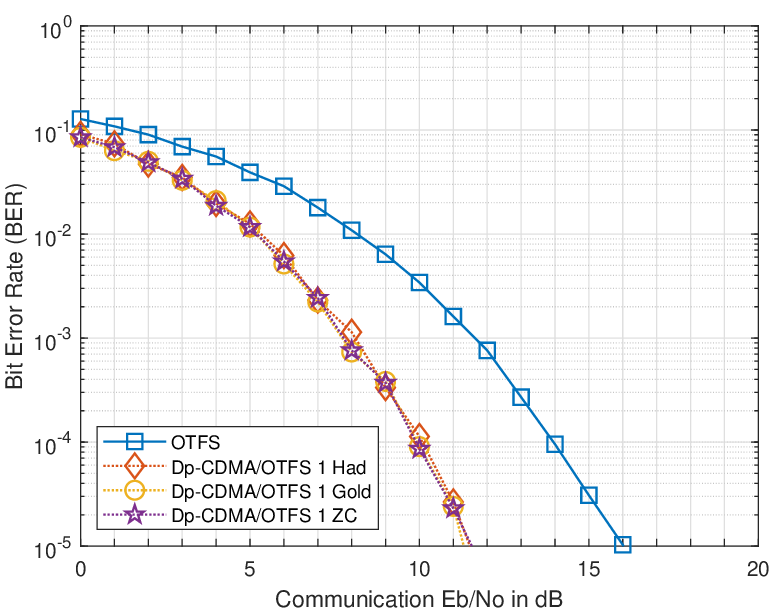}
        \caption{\acs*{otfs} and \acs*{dp-cdma-otfs}}
        \label{fig:ber_n_mult_1_dop_only}
    \end{subfigure}
    \begin{subfigure}{.45\textwidth}
        \includegraphics[width=0.9\linewidth]{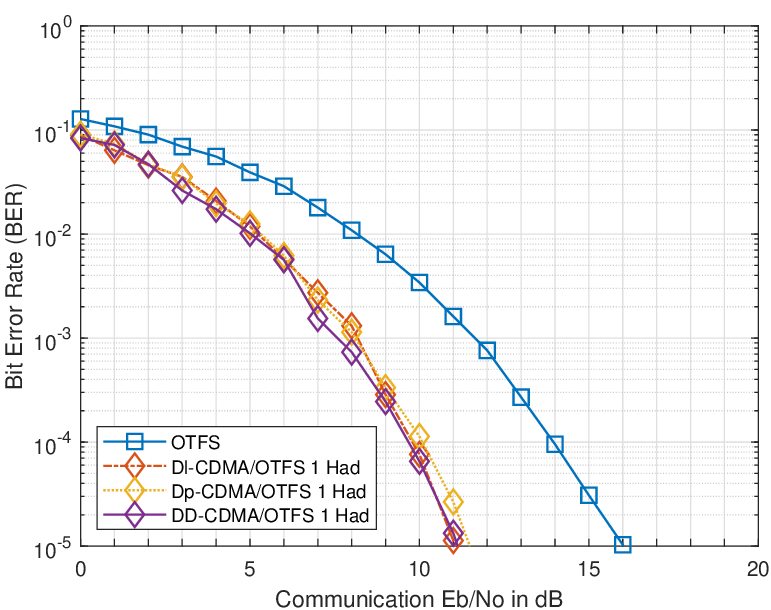}
        \caption{\acs*{otfs} and Hadamard sequence}
        \label{fig:ber_n_mult_1_had_only}
    \end{subfigure}
    \begin{subfigure}{.45\textwidth}
        \includegraphics[width=0.9\linewidth]{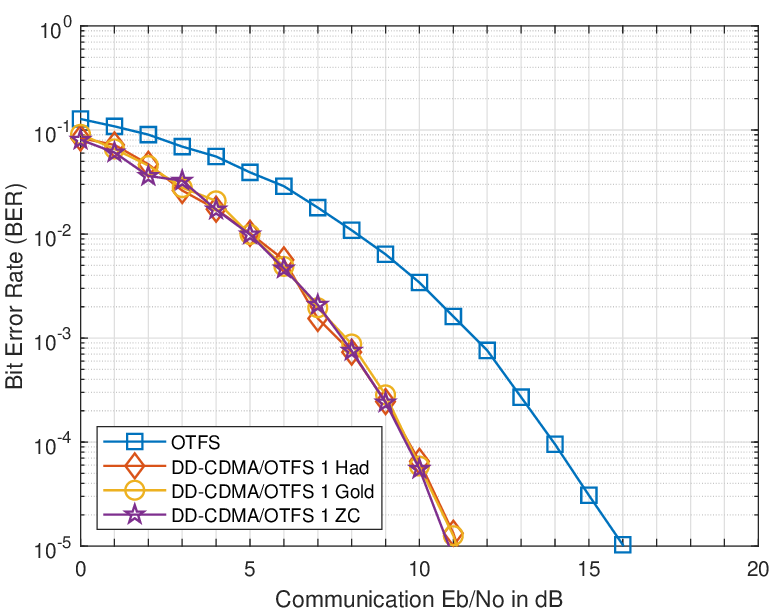}
        \caption{\acs*{otfs} and \acs*{dd-cdma-otfs}}
        \label{fig:ber_n_mult_1_dd_only}
    \end{subfigure}
    \begin{subfigure}{.45\textwidth}
        \includegraphics[width=0.9\linewidth]{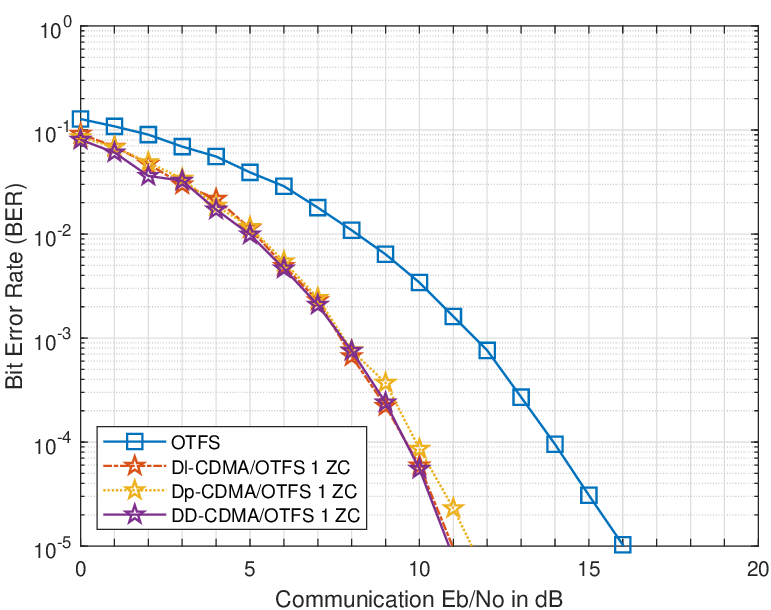}
        \caption{\acs*{otfs} and Zadoff-Chu sequence}
        \label{fig:ber_n_mult_1_zc_only}
    \end{subfigure}
    \caption{\acs*{ber} of \acs*{qpsk} Gold, Hadamard, and Zadoff-Chu sequence spreading for $N_{mult} = 1$ for \acs*{dl-cdma-otfs}, \acs*{dp-cdma-otfs}, and \acs*{dd-cdma-otfs}, and of \acs*{otfs} \acs*{qpsk}}
    \label{fig:ber_n_mult_1}
\end{figure*}

\begin{figure}[tp]
    \centering
    \includegraphics[width=.45\textwidth]{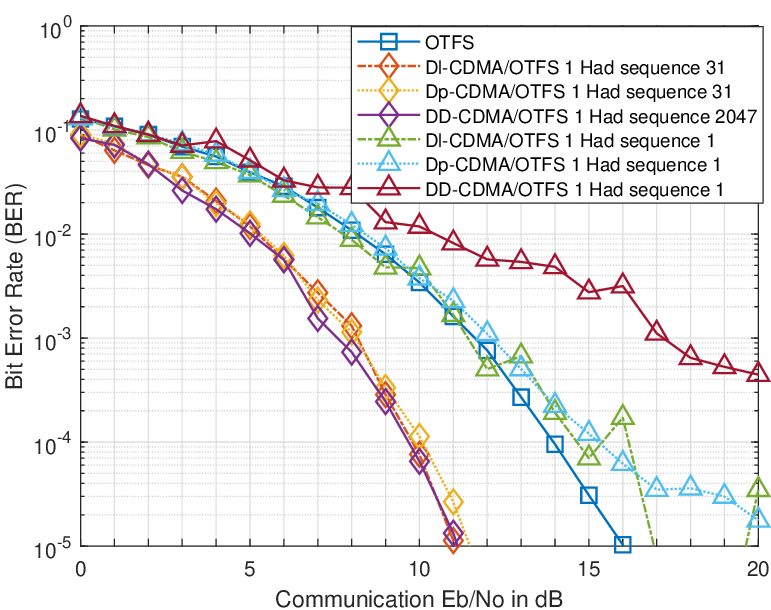}
    \caption{\acs*{ber} of \acs*{qpsk} Hadamard sequences $1$ and $31$ spreading for $N_{mult} = 1$ for \acs*{dl-cdma-otfs} and \acs*{dp-cdma-otfs}, Hadamard sequences $1$ and $2047$ spreading for \acs*{dd-cdma-otfs}, and of \acs*{otfs} \acs*{qpsk}}
    \label{fig:ber_n_mult_1_had_comp}
\end{figure}

\begin{figure*}[tp]
    \centering
    \begin{subfigure}{.45\textwidth}
        \includegraphics[width=0.9\linewidth]{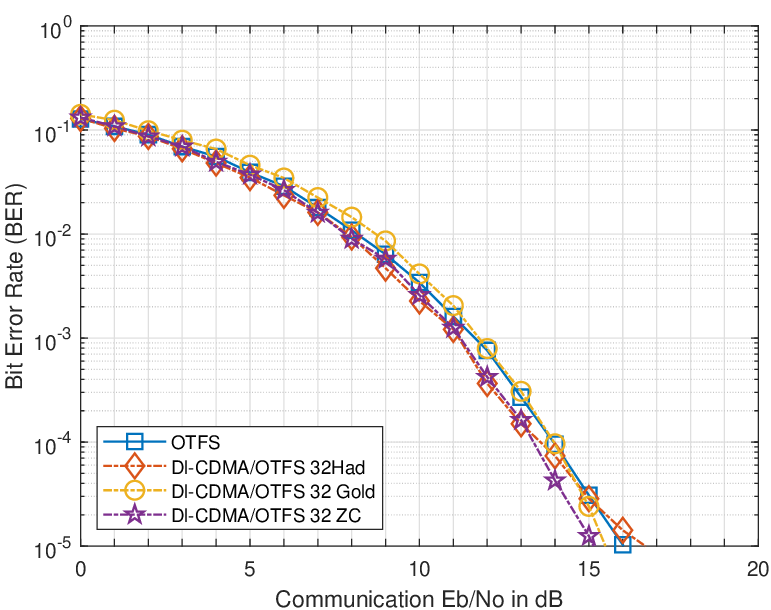}
        \caption{\acs*{otfs} and \acs*{dl-cdma-otfs}}
        \label{fig:ber_n_mult_half_max_del_only}
    \end{subfigure}
    \begin{subfigure}{.45\textwidth}
        \includegraphics[width=0.9\linewidth]{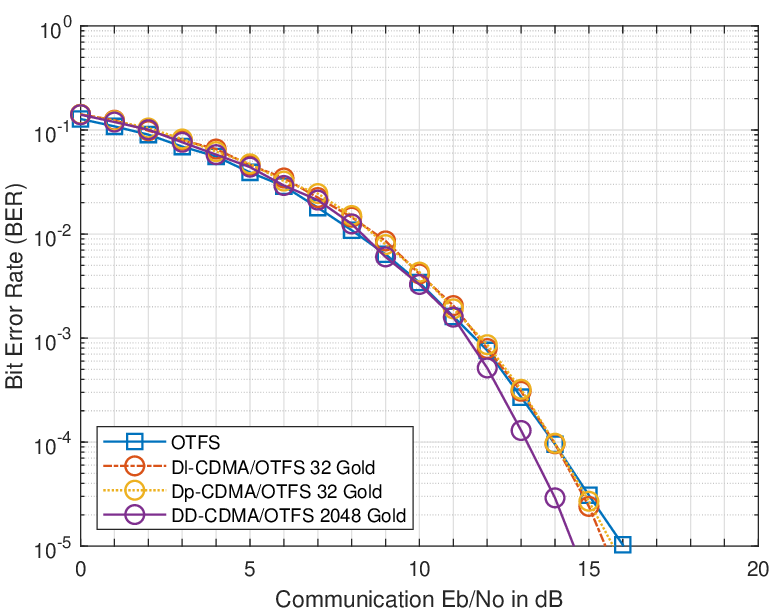}
        \caption{\acs*{otfs} and Gold sequence}
        \label{fig:ber_n_mult_half_max_gold_only}
    \end{subfigure}
    \begin{subfigure}{.45\textwidth}
        \includegraphics[width=0.9\linewidth]{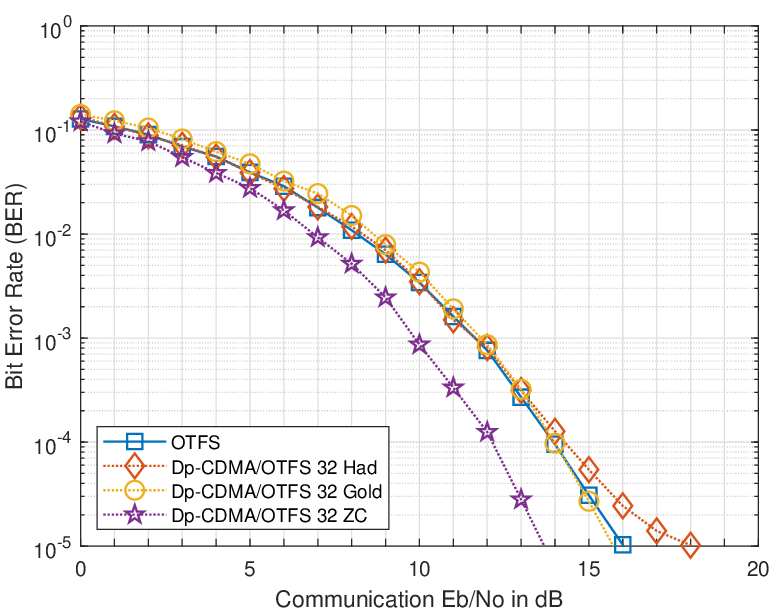}
        \caption{\acs*{otfs} and \acs*{dp-cdma-otfs}}
        \label{fig:ber_n_mult_half_max_dop_only}
    \end{subfigure}
    \begin{subfigure}{.45\textwidth}
        \includegraphics[width=0.9\linewidth]{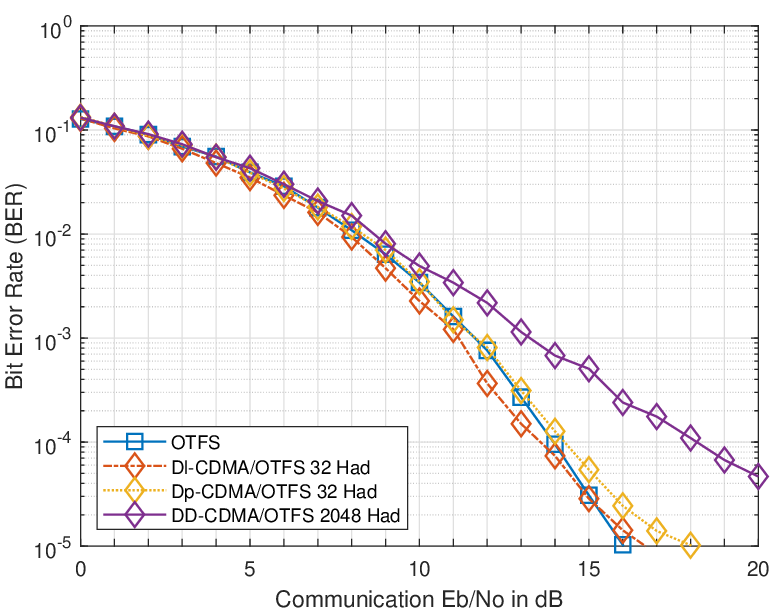}
        \caption{\acs*{otfs} and Hadamard sequence}
        \label{fig:ber_n_mult_half_max_had_only}
    \end{subfigure}
    \begin{subfigure}{.45\textwidth}
        \includegraphics[width=0.9\linewidth]{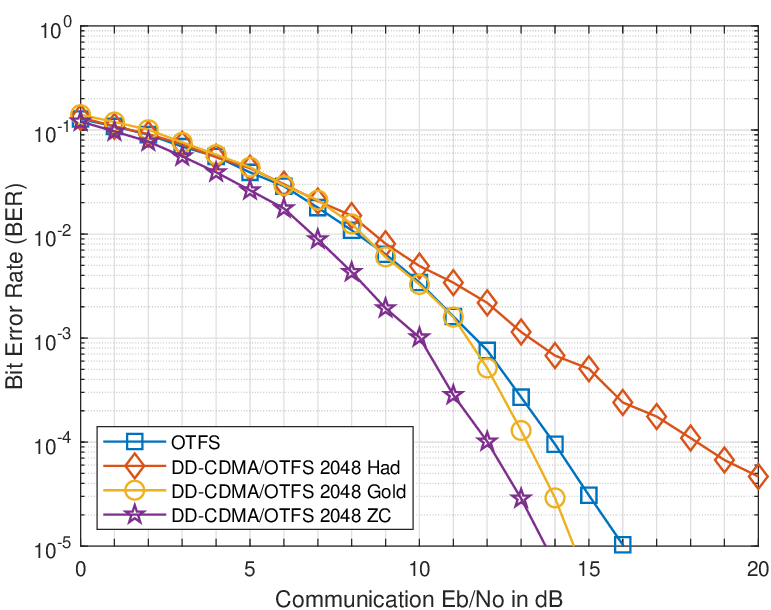}
        \caption{\acs*{otfs} and \acs*{dd-cdma-otfs}}
        \label{fig:ber_n_mult_half_max_dd_only}
    \end{subfigure}
    \begin{subfigure}{.45\textwidth}
        \includegraphics[width=0.9\linewidth]{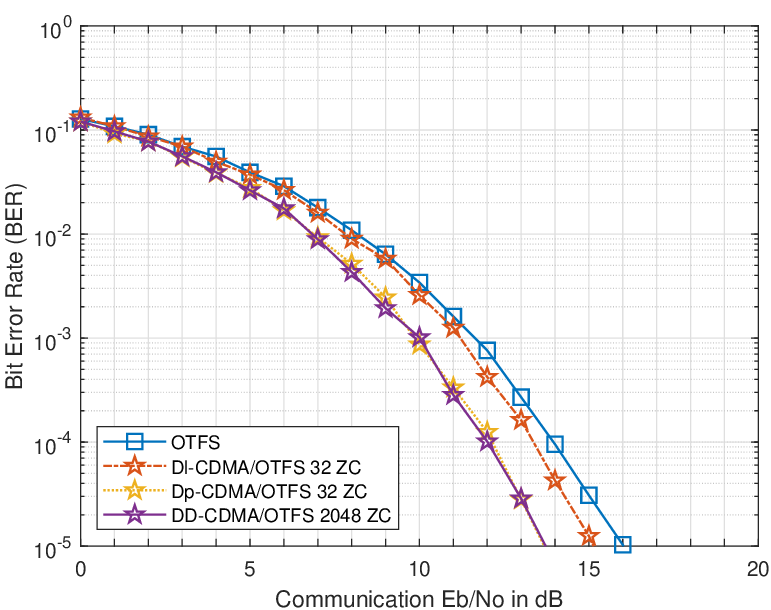}
        \caption{\acs*{otfs} and Zadoff-Chu sequence}
        \label{fig:ber_n_mult_half_max_zc_only}
    \end{subfigure}
    \caption{\acs*{ber} of \acs*{qpsk} Gold, Hadamard, and Zadoff-Chu sequence spreading for $N_{mult} = 32$ for \acs*{dl-cdma-otfs} and \acs*{dp-cdma-otfs}, $N_{mult} = 2048$ for \acs*{dd-cdma-otfs}, and of \acs*{otfs} \acs*{qpsk}}
    \label{fig:ber_n_mult_half_max}
\end{figure*}

\begin{figure*}[tp]
    \centering
    \begin{subfigure}{.45\textwidth}
        \includegraphics[width=0.9\linewidth]{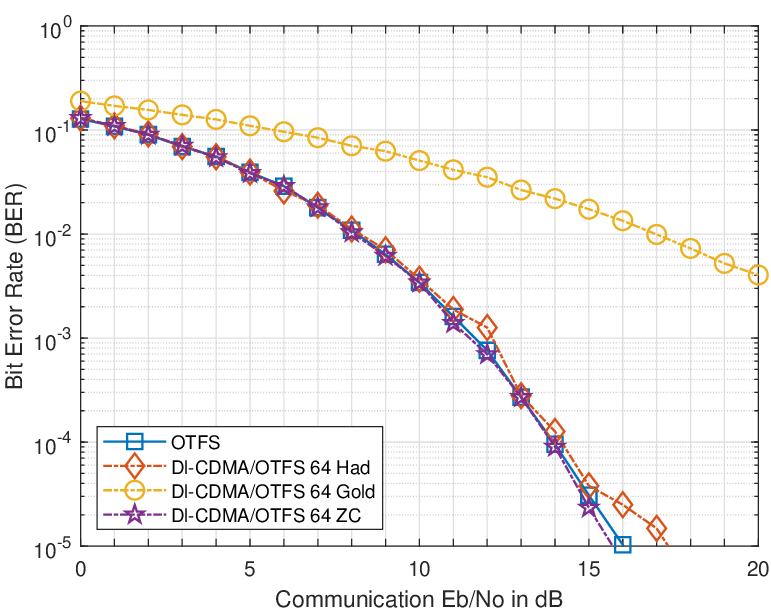}
        \caption{\acs*{otfs} and \acs*{dl-cdma-otfs}}
        \label{fig:ber_n_mult_max_del_only}
    \end{subfigure}
    \begin{subfigure}{.45\textwidth}
        \includegraphics[width=0.9\linewidth]{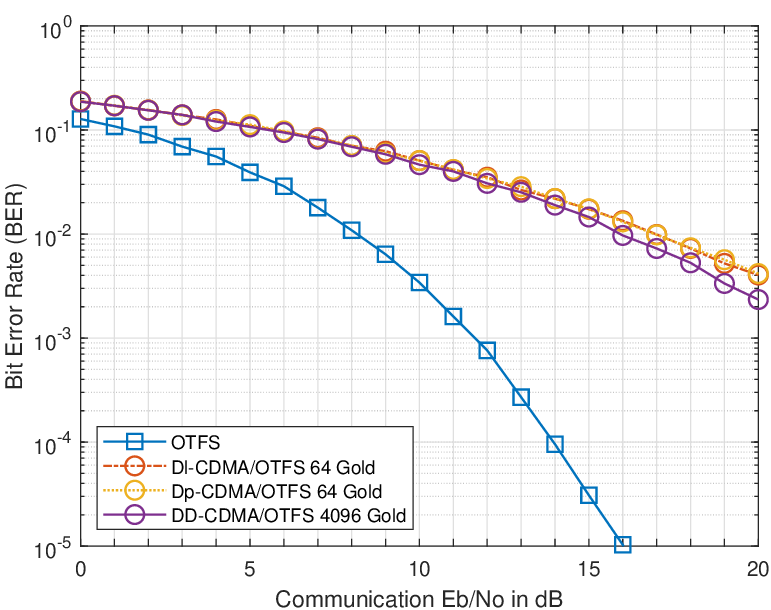}
        \caption{\acs*{otfs} and Gold sequence}
        \label{fig:ber_n_mult_max_gold_only}
    \end{subfigure}
    \begin{subfigure}{.45\textwidth}
        \includegraphics[width=0.9\linewidth]{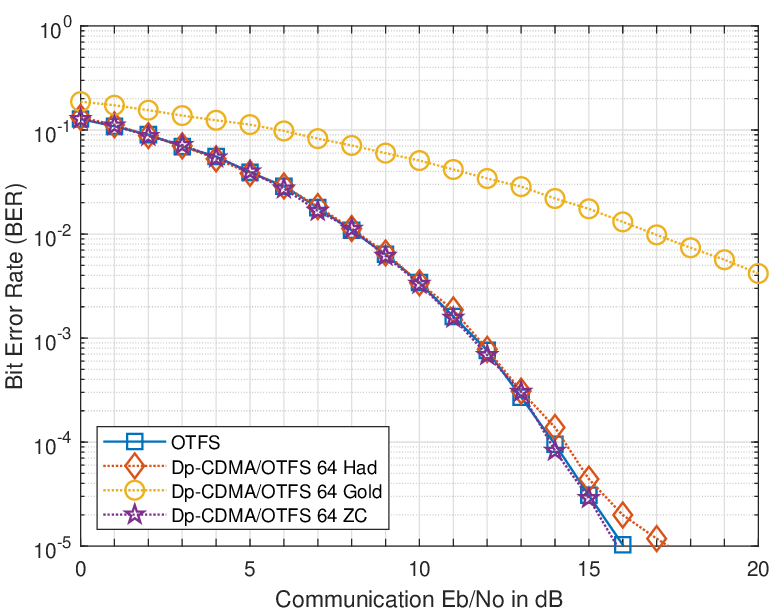}
        \caption{\acs*{otfs} and \acs*{dp-cdma-otfs}}
        \label{fig:ber_n_mult_max_dop_only}
    \end{subfigure}
    \begin{subfigure}{.45\textwidth}
        \includegraphics[width=0.9\linewidth]{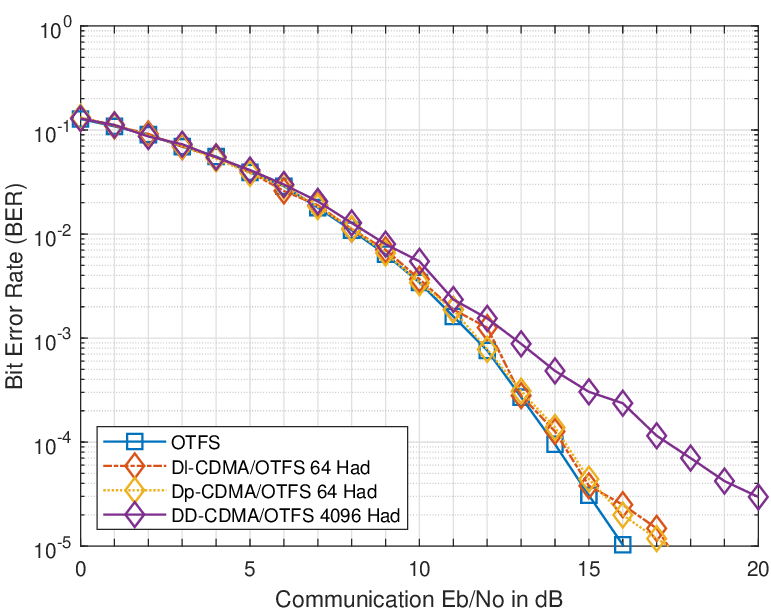}
        \caption{\acs*{otfs} and Hadamard sequence}
        \label{fig:ber_n_mult_max_had_only}
    \end{subfigure}
    \begin{subfigure}{.45\textwidth}
        \includegraphics[width=0.9\linewidth]{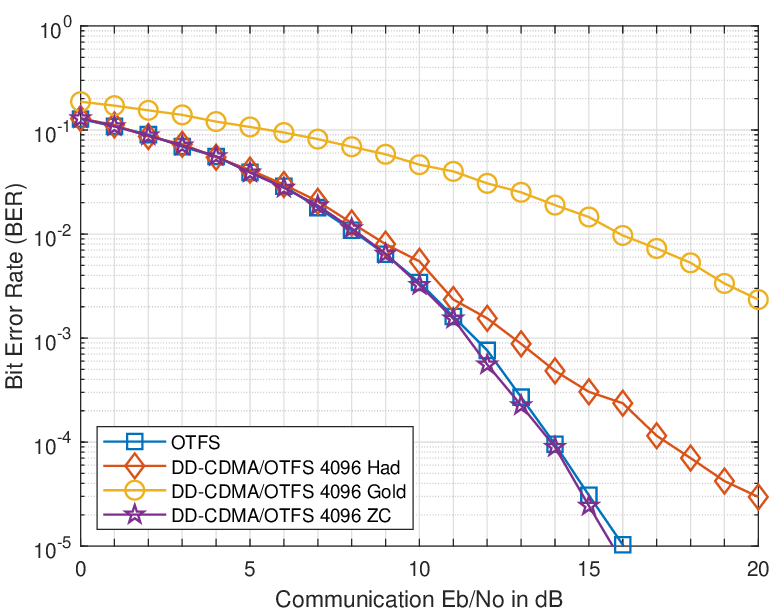}
        \caption{\acs*{otfs} and \acs*{dd-cdma-otfs}}
        \label{fig:ber_n_mult_max_dd_only}
    \end{subfigure}
    \begin{subfigure}{.45\textwidth}
        \includegraphics[width=0.9\linewidth]{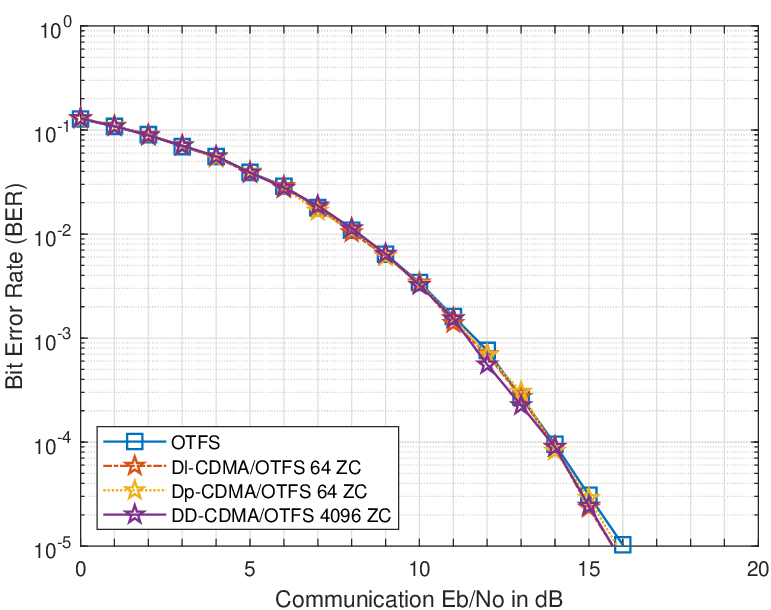}
        \caption{\acs*{otfs} and Zadoff-Chu sequence}
        \label{fig:ber_n_mult_max_zc_only}
    \end{subfigure}
    \caption{\acs*{ber} of \acs*{qpsk} Gold, Hadamard, and Zadoff-Chu sequence spreading for $N_{mult} = 64$ for \acs*{dl-cdma-otfs} and \acs*{dp-cdma-otfs}, $N_{mult} = 4096$ for \acs*{dd-cdma-otfs}, and of \acs*{otfs} \acs*{qpsk}}
    \label{fig:ber_n_mult_max}
\end{figure*}

The simulation parameters used are shown in Table \ref{tab:com_sim_vars}. The \ac{ber} of \ac{qpsk} Gold, Hadamard, and Zadoff-Chu sequence spreading for \ac{dl-cdma-otfs}, \ac{dp-cdma-otfs}, and \ac{dd-cdma-otfs} is portrayed in Figure \ref{fig:ber_n_mult_1} for $N_{mult} = 1$. The \ac{ber} of \ac{otfs} \ac{qpsk} is also shown in Figure \ref{fig:ber_n_mult_1}. At the minimum throughput, all system configurations relying on the three sequences have similar \ac{ber} performances. The \ac{ber} of \ac{cdma-otfs} for $N_{mult} = 1$ is much lower than that of \ac{otfs}, albeit at a large cost of throughput, which is reduced by a factor of $64$ for \ac{dl-cdma-otfs} and \ac{dp-cdma-otfs}, and a factor of $4096$ for \ac{dd-cdma-otfs}.

The performance of Hadamard sequences for $N_{mult} = 1$ depends on the specific Hadamard sequence selected, as shown in Figure \ref{fig:ber_n_mult_1_had_comp}. When the first Hadamard sequence of the Hadamard matrix is selected (Hadamard sequence 1), the \ac{ber} for \ac{cdma-otfs} is higher than that of \ac{otfs}, as this sequence is a vector of $1$.

The \ac{ber} of \ac{qpsk} Gold, Hadamard, and Zadoff-Chu sequence spreading for $N_{mult} = 32$ for \ac{dl-cdma-otfs} and \ac{dp-cdma-otfs}, $N_{mult} = 2048$ for \ac{dd-cdma-otfs}, and of \ac{otfs} \ac{qpsk} is shown in Figure \ref{fig:ber_n_mult_half_max} for a throughput of 1 \ac{bpcu} for \ac{cdma-otfs}. For 1 \ac{bpcu}, the Zadoff-Chu sequence outperforms \ac{otfs} for all spreading configurations, as Zadoff-Chu sequences are resistant to delay and Doppler interference. The \ac{ber} of Gold sequence spreading is similar to that of \ac{otfs} for all spreading configurations, with the \ac{ber} of Gold \ac{dd-cdma-otfs} slightly lower than that of \ac{otfs} at high \ac{eb/no}. This is because the length of the Gold sequences for \ac{dd-cdma-otfs} is longer than for \ac{dl-cdma-otfs} and \ac{dp-cdma-otfs}. The \ac{ber} performance of Hadamard \ac{dl-cdma-otfs} and \ac{dp-cdma-otfs} is similar to \ac{otfs} for the majority of the \ac{eb/no} range, but its \ac{ber} is slightly higher at larger \ac{eb/no}, as Hadamard sequences are vulnerable to multi-path interference. The \ac{ber} of Hadamard \ac{dd-cdma-otfs} is higher than that of \ac{otfs}, since the spreading sequence experiences multi-path interference in both the delay and Doppler domains.

The \ac{ber} of \ac{qpsk} Gold, Hadamard, and Zadoff-Chu sequence spreading for $N_{mult} = 64$ for \ac{dl-cdma-otfs} and \ac{dp-cdma-otfs}, $N_{mult} = 4096$ for \ac{dd-cdma-otfs}, and of \ac{otfs} \ac{qpsk} is shown in Figure \ref{fig:ber_n_mult_max}, for a throughput of 2 \ac{bpcu}. At 2 \ac{bpcu}, the performance of Gold sequences is poor, as their lack of orthogonality leads to increased inter-symbol interference. The \ac{ber} of Zadoff-Chu sequences is almost identical to \ac{otfs}. The \ac{ber} for 2 \ac{bpcu} is higher than for a throughput of 1 \ac{bpcu}, as the sequences experience a small amount of inter-symbol interference, despite their orthogonality. The performance of Hadamard sequences is similar to the half throughput case  of 1 \ac{bpcu}, as the sequences experience little inter-symbol interference due to their orthogonality to each other.

\subsection{Sensing \acs*{rmse} Results}
\label{sec:results_rmse}

\begin{table}[tp]
    \centering
    {
    \caption{Sensing simulation parameters}
    \begin{tabular}{|l|r|}
        \hline
        \multicolumn{1}{|c|}{Variable} & \multicolumn{1}{|c|}{Value}\\
        \hhline{|=|=|}
        Carrier frequency $f_{c}$ & $40$ GHz\\
        \hline
        Subcarrier spacing $\Delta f$ & $120$ kHz\\
        \hline
        Number of sensing targets $P_{t}$ & $1$\\
        \hline
        Rician K factor $\kappa_{sen}$ & $10$ dB\\
        \hline
        Number of subcarriers $M$ & $64$\\
        \hline
        Number of symbols sent per frame $N$ & $64$\\
        \hline
        Number of frames simulated per \acs*{eb/no} & $4000$\\
        \hline
    \end{tabular}
    \label{tab:sen_sim_vars}
    }
\end{table}

\begin{figure*}[tp]
    \centering
    \begin{subfigure}{.45\textwidth}
        \includegraphics[width=0.9\linewidth]{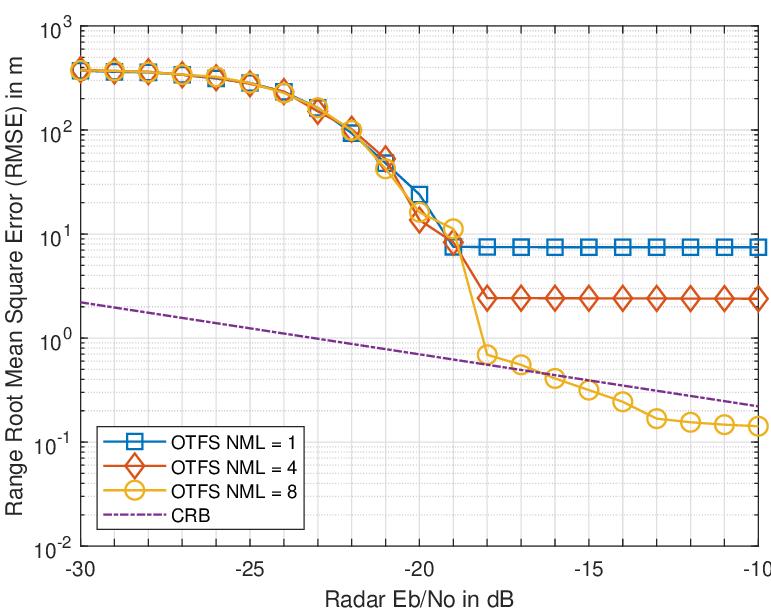}
        \caption{Range \acs*{rmse}}
        \label{fig:r_rmse_otfs_nml_1_4_8_p_1}
    \end{subfigure}
    \begin{subfigure}{.45\textwidth}
        \includegraphics[width=0.9\linewidth]{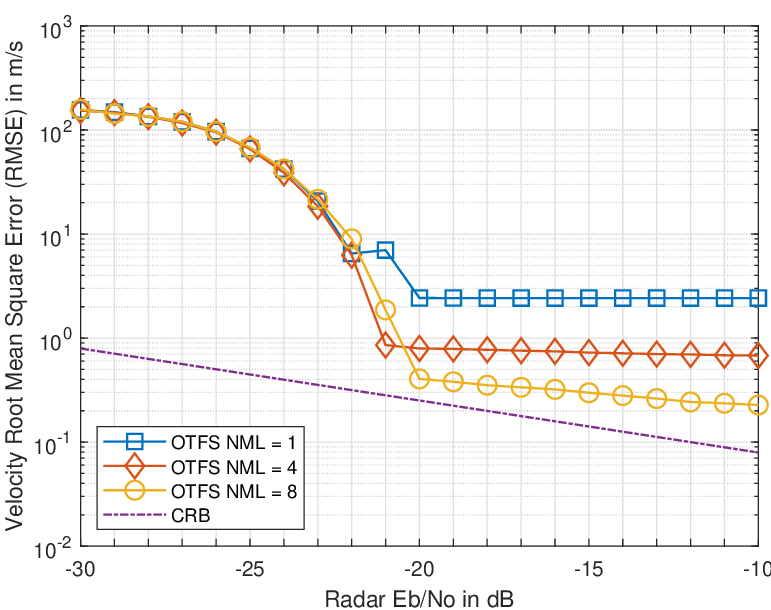}
        \caption{Velocity \acs*{rmse}}
        \label{fig:v_rmse_otfs_nml_1_4_8_p_1}
    \end{subfigure}
    \caption{Range and velocity \acs*{rmse} of \acs*{otfs} \acs*{qpsk} for $N_{ML} = 1$, $4$, and $8$, $R_{t} = 500$ m, $V_{t} = 200$ m/s, and $P_{n} = 0$ \acs*{nlos} paths}
    \label{fig:rmse_otfs_nml_1_4_8_p_1}
\end{figure*}

\begin{figure*}[tp]
    \centering
    \begin{subfigure}{.45\textwidth}
        \includegraphics[width=0.9\linewidth]{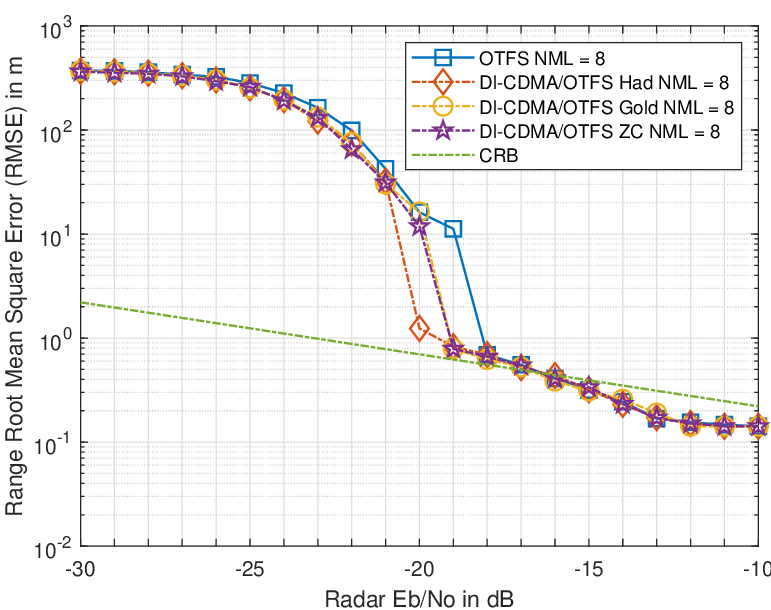}
        \caption{\acs*{otfs} and \acs*{dl-cdma-otfs}}
        \label{fig:r_rmse_n_mult_max_nml_8_del_only_p_1}
    \end{subfigure}
    \begin{subfigure}{.45\textwidth}
        \includegraphics[width=0.9\linewidth]{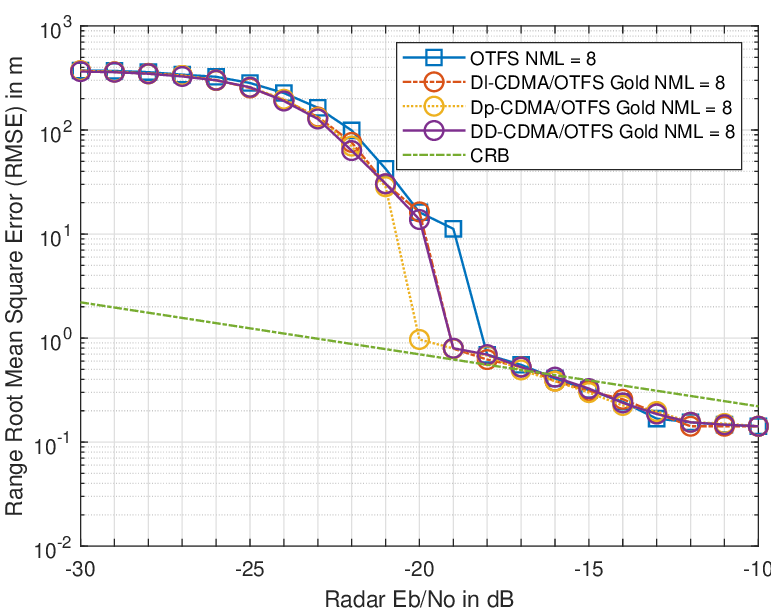}
        \caption{\acs*{otfs} and Gold sequence}
        \label{fig:r_rmse_n_mult_max_nml_8_gold_only_p_1}
    \end{subfigure}
    \begin{subfigure}{.45\textwidth}
        \includegraphics[width=0.9\linewidth]{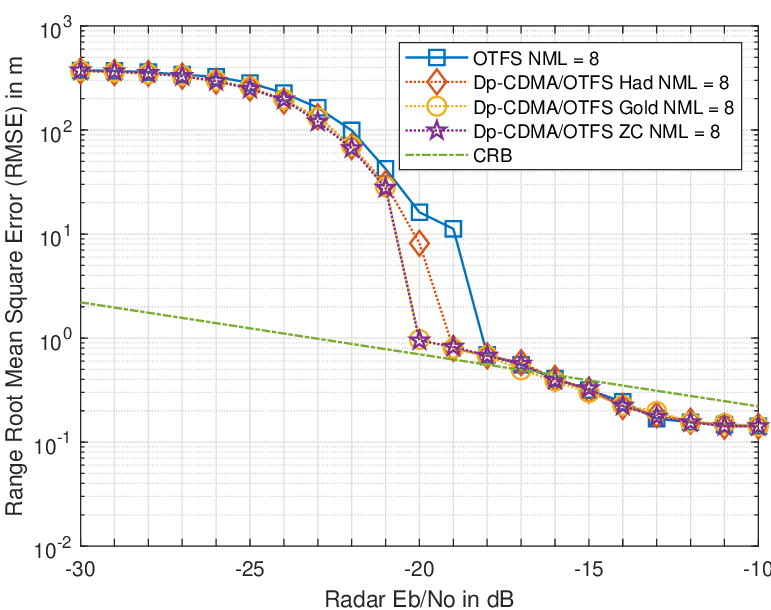}
        \caption{\acs*{otfs} and \acs*{dp-cdma-otfs}}
        \label{fig:r_rmse_n_mult_max_nml_8_dop_only_p_1}
    \end{subfigure}
    \begin{subfigure}{.45\textwidth}
        \includegraphics[width=0.9\linewidth]{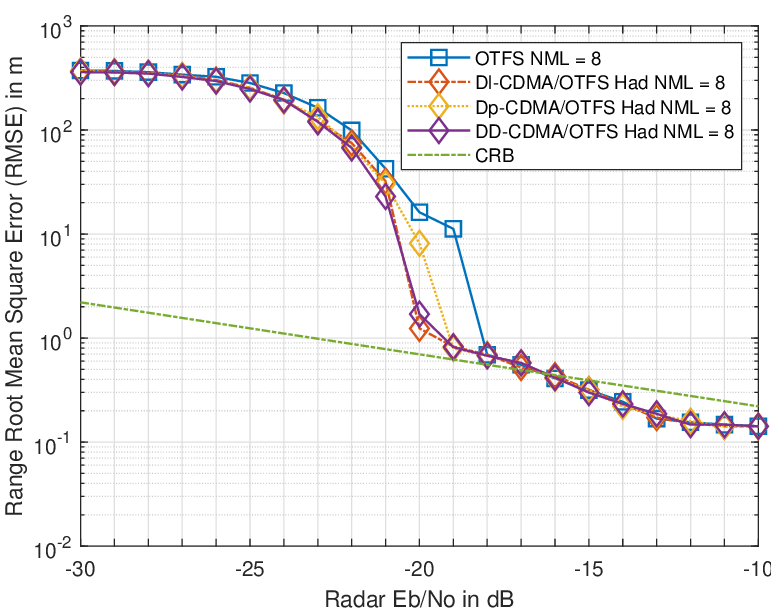}
        \caption{\acs*{otfs} and Hadamard sequence}
        \label{fig:r_rmser_n_mult_max_nml_8_had_only_p_1}
    \end{subfigure}
    \begin{subfigure}{.45\textwidth}
        \includegraphics[width=0.9\linewidth]{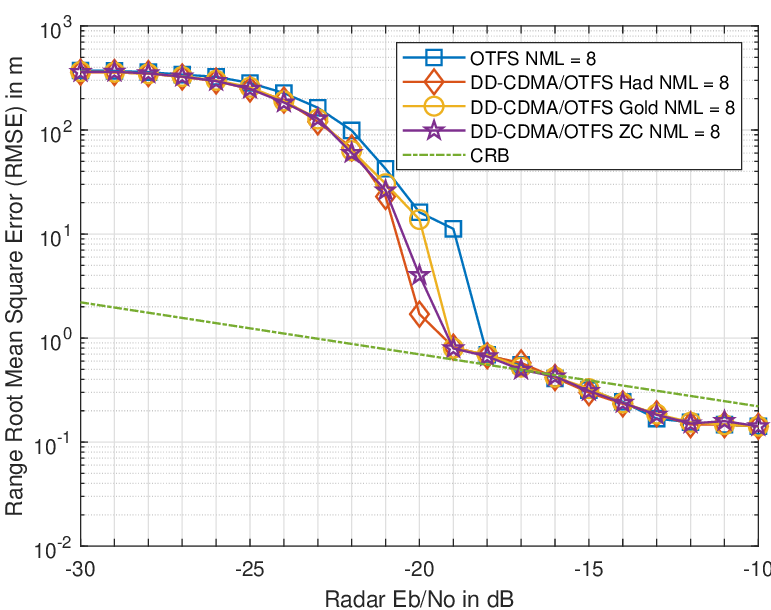}
        \caption{\acs*{otfs} and \acs*{dd-cdma-otfs}}
        \label{fig:r_rmse_n_mult_max_nml_8_dd_only_p_1}
    \end{subfigure}
    \begin{subfigure}{.45\textwidth}
        \includegraphics[width=0.9\linewidth]{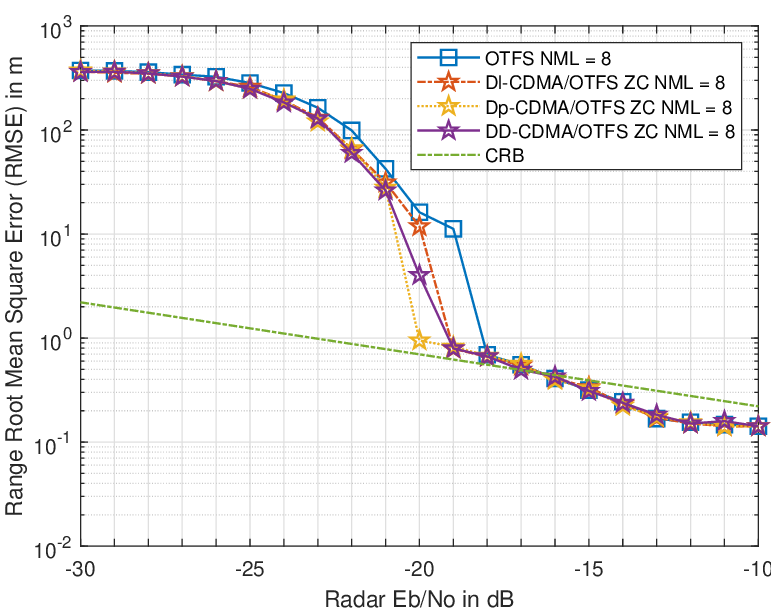}
        \caption{\acs*{otfs} and Zadoff-Chu sequence}
        \label{fig:r_rmse_n_mult_max_nml_8_zc_only_p_1}
    \end{subfigure}
    \caption{Range \acs*{rmse} of \acs*{qpsk} Gold, Hadamard, and Zadoff-Chu sequence spreading for $N_{mult} = 64$ for \acs*{dl-cdma-otfs} and \acs*{dp-cdma-otfs}, $N_{mult} = 4096$ for \acs*{dd-cdma-otfs}, and of \acs*{otfs} \acs*{qpsk} for $N_{ML} = 8$, $R_{t} = 500$ m, $V_{t} = 200$ m/s, and $P_{n} = 0$ \acs*{nlos} paths}
    \label{fig:r_rmse_n_mult_max_nml_8_p_1}
\end{figure*}

\begin{figure*}[tp]
    \centering
    \begin{subfigure}{.45\textwidth}
        \includegraphics[width=0.9\linewidth]{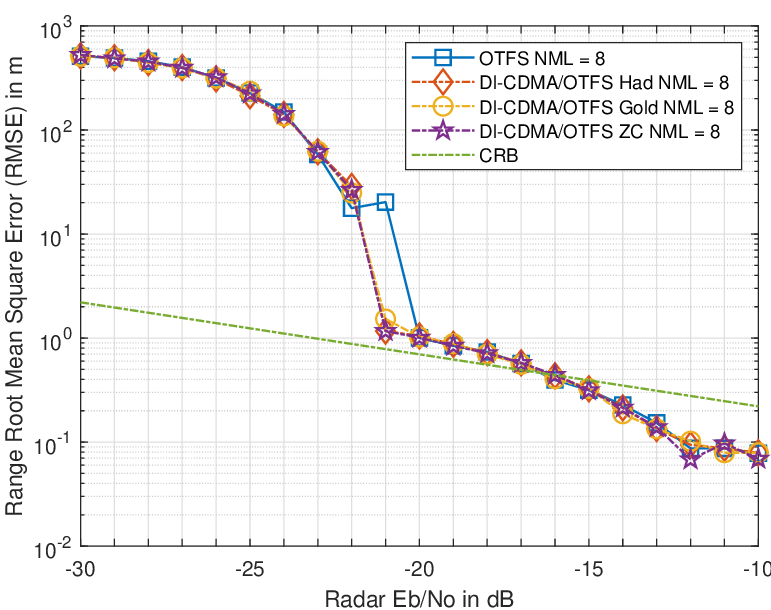}
        \caption{\acs*{otfs} and \acs*{dl-cdma-otfs}}
        \label{fig:r_rmse_n_mult_max_nml_8_del_only_p_8_k_10}
    \end{subfigure}
    \begin{subfigure}{.45\textwidth}
        \includegraphics[width=0.9\linewidth]{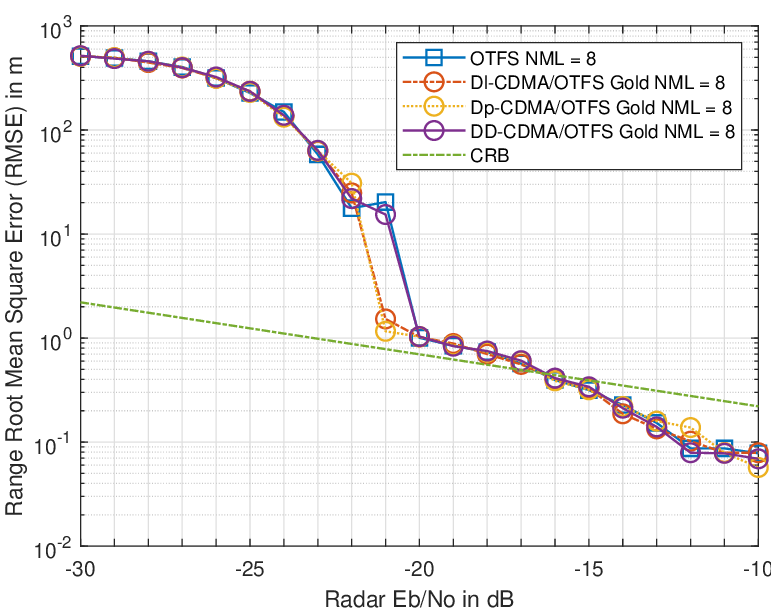}
        \caption{\acs*{otfs} and Gold sequence}
        \label{fig:r_rmse_n_mult_max_nml_8_gold_only_p_8_k_10}
    \end{subfigure}
    \begin{subfigure}{.45\textwidth}
        \includegraphics[width=0.9\linewidth]{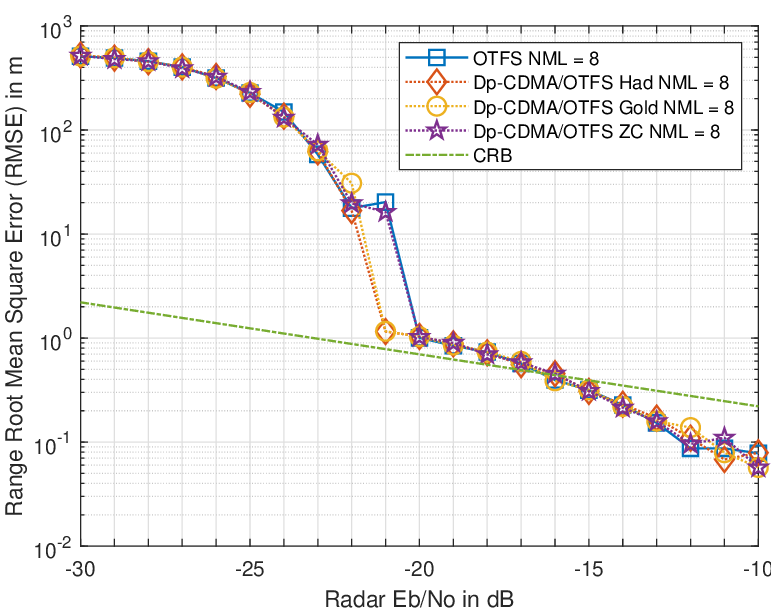}
        \caption{\acs*{otfs} and \acs*{dp-cdma-otfs}}
        \label{fig:r_rmse_n_mult_max_nml_8_dop_only_p_8_k_10}
    \end{subfigure}
    \begin{subfigure}{.45\textwidth}
        \includegraphics[width=0.9\linewidth]{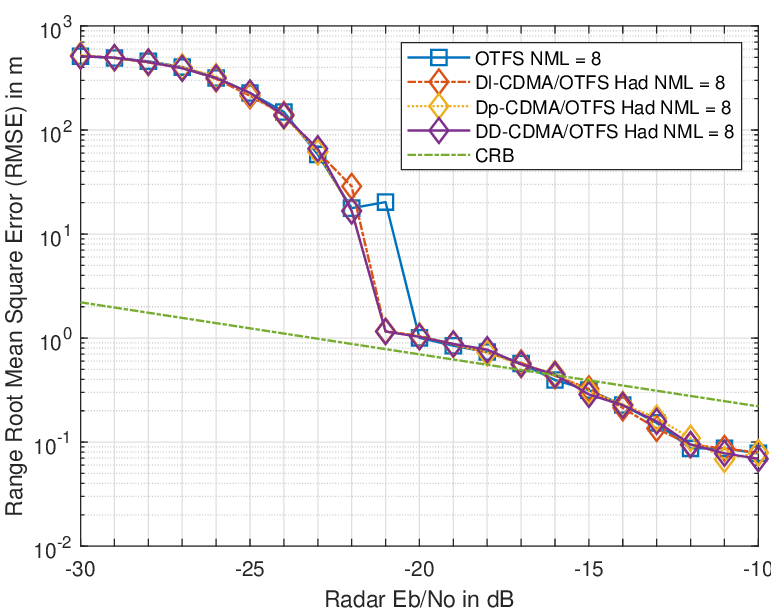}
        \caption{\acs*{otfs} and Hadamard sequence}
        \label{fig:r_rmser_n_mult_max_nml_8_had_only_p_8_k_10}
    \end{subfigure}
    \begin{subfigure}{.45\textwidth}
        \includegraphics[width=0.9\linewidth]{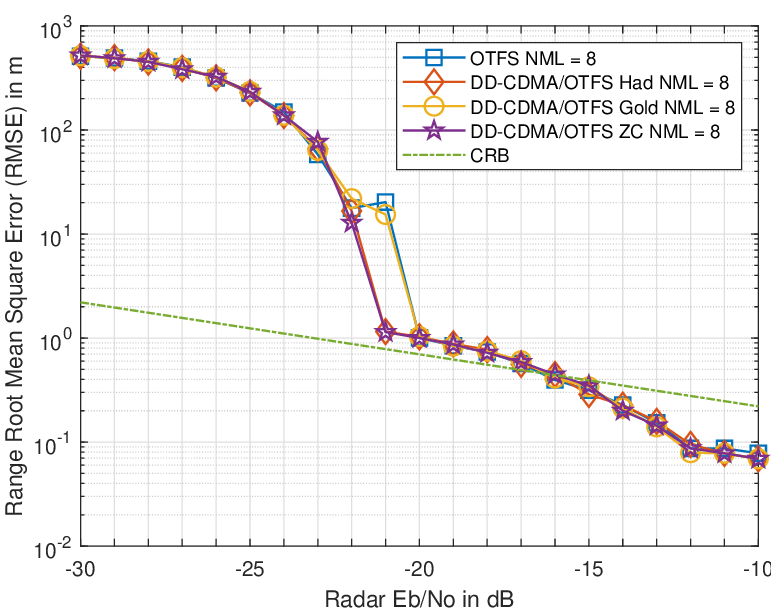}
        \caption{\acs*{otfs} and \acs*{dd-cdma-otfs}}
        \label{fig:r_rmse_n_mult_max_nml_8_dd_only_p_8_k_10}
    \end{subfigure}
    \begin{subfigure}{.45\textwidth}
        \includegraphics[width=0.9\linewidth]{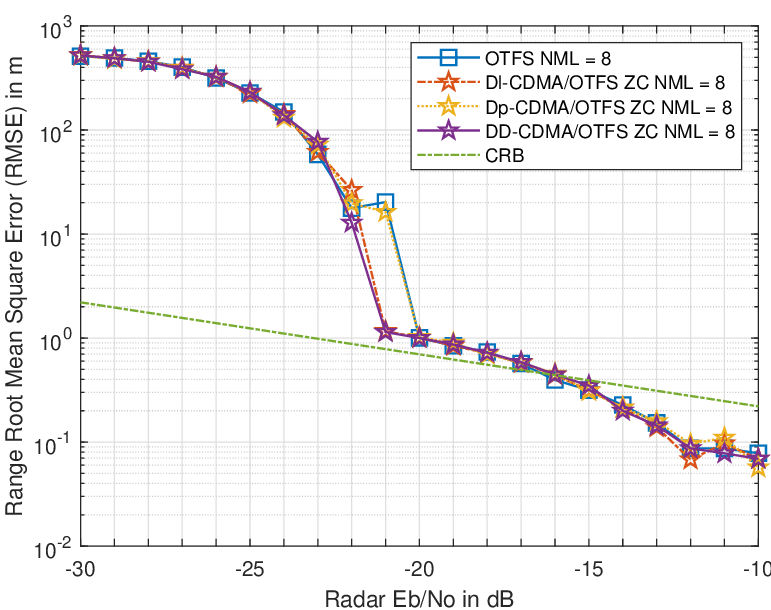}
        \caption{\acs*{otfs} and Zadoff-Chu sequence}
        \label{fig:r_rmse_n_mult_max_nml_8_zc_only_p_8_k_10}
    \end{subfigure}
    \caption{Range \acs*{rmse} of \acs*{qpsk} Gold, Hadamard, and Zadoff-Chu sequence spreading for $N_{mult} = 64$ for \acs*{dl-cdma-otfs} and \acs*{dp-cdma-otfs}, $N_{mult} = 4096$ for \acs*{dd-cdma-otfs}, and of \acs*{otfs} \acs*{qpsk} for $N_{ML} = 8$, $R_{t} = 200$ m, $V_{t} = 110$ m/s, and $P_{n} = 7$ \acs*{nlos} paths}
    \label{fig:r_rmse_n_mult_max_nml_8_p_8_k_10}
\end{figure*}

The simulation parameters used are shown in Table \ref{tab:sen_sim_vars}. The range and velocity \ac{rmse} of \ac{qpsk} \ac{otfs} are shown in Figure \ref{fig:rmse_otfs_nml_1_4_8_p_1} for $N_{ML} = 1$, $4$, and $8$, $R_{t} = 500$ m, $V_{t} = 200$ m/s, and $P_{n} = 0$ \ac{nlos} paths. The \ac{rmse} is dominated by the integer index estimation at \ac{eb/no} below $-18$ dB, but by the fractional index estimation at higher \ac{eb/no}. As $N_{ML}$ is increased, the system resolution is increased, which leads to a potentially lower error floor, as seen in Figure \ref{fig:r_rmse_otfs_nml_1_4_8_p_1}.

In Figure \ref{fig:v_rmse_otfs_nml_1_4_8_p_1}, the closest fractional index to the target velocity is unchanged for $N_{ML} = 4$ and $8$, hence the error floor is identical. As $N_{ML}$ is increased, the fractional index detection algorithm becomes more sensitive to \ac{awgn}, which leads to the \ac{rmse} of the estimate reaching the error floor at higher \ac{eb/no} than for $N_{ML} = 4$. All subsequent results use $N_{ML} = 8$.

In Figure \ref{fig:v_rmse_otfs_nml_1_4_8_p_1}, the error floor is higher than the \ac{crb}, but in Figure \ref{fig:r_rmse_otfs_nml_1_4_8_p_1}, the error floor is lower than the \ac{crb} for $N_{ML} = 8$. This is due to the discrete properties of the estimator. The target parameter estimator outputs a value from a discrete set, whereas the \ac{crb} assumes a continuous set. Hence, when the discrete estimator output is close to the target parameter value, the sensing error can be smaller than the \ac{crb} in the \ac{eb/no} range considered. At higher \ac{eb/no}, the \ac{crb} will reduce further, below the error floor.

As similar trends are observed for range and velocity estimation, only the range \ac{rmse} results will be presented.

The range \ac{rmse} of \ac{qpsk} Gold, Hadamard, and Zadoff-Chu sequence spreading are shown in Figure \ref{fig:r_rmse_n_mult_max_nml_8_p_1} for $N_{mult} = 64$ for \ac{dl-cdma-otfs} and \ac{dp-cdma-otfs}, $N_{mult} = 4096$ for \ac{dd-cdma-otfs}, and of \ac{otfs} \ac{qpsk} for $N_{ML} = 8$, $R_{t} = 500$ m, $V_{t} = 200$ m/s, and $P_{n} = 0$ \ac{nlos} paths.

For most of the \ac{cdma-otfs} systems, the \ac{eb/no} at which the \ac{rmse} is dominated by the fractional index estimation error is 1 dB lower than for \ac{otfs}, except for Hadamard \ac{dl-cdma-otfs}, Gold and Zadoff-Chu \ac{dp-cdma-otfs}, where it is 2dB lower than for \ac{otfs}. This is because the spreading codes distribute the signal over multiple indices, thereby creating a more uniform transmitted signal compared to \ac{otfs}. It has been shown \cite{on_fundamental_limits_isac} that more deterministic signals lead to a superior sensing performance.

The range \ac{rmse} of \ac{qpsk} Gold, Hadamard, and Zadoff-Chu sequence spreading are shown in Figure \ref{fig:r_rmse_n_mult_max_nml_8_p_8_k_10} for $N_{mult} = 64$ for \ac{dl-cdma-otfs} and \ac{dp-cdma-otfs}, $N_{mult} = 4096$ for \ac{dd-cdma-otfs}, and of \ac{otfs} \ac{qpsk} for $N_{ML} = 8$, $R_{t} = 200$ m, $V_{t} = 110$ m/s, and $P_{n} = 7$ \ac{nlos} paths.

For most of the \ac{cdma-otfs} systems, the \ac{eb/no} at which the \ac{rmse} is dominated by the fractional index estimation error is 1 dB lower than for \ac{otfs}, except for Zadoff-Chu \ac{dp-cdma-otfs} and Gold \ac{dd-cdma-otfs}, where it is at the same \ac{eb/no} as \ac{otfs}. As the velocity is decreased, the sensing performance advantage of \ac{cdma-otfs} is slightly decreased, as the distortion caused by the Doppler shifts of the propagation paths is reduced.

\section{Conclusions}
\label{sec:conc}

\begin{table}[tp]
    \centering
    {
    \caption{\acs*{eb/no} at which a \acs*{ber} of $10^{-4}$ is reached for \acs*{cdma-otfs} at half load relative to \acs*{otfs}}
    \begin{tabular}{|c|c|c|c|}
        \hline
        \backslashbox{Configuration}{Sequence} & Gold & Hadamard & Zadoff-Chu\\
        \hline
        \acs*{dl-cdma-otfs} & $0$ dB & $0$ dB & $-0.5$ dB\\
        \hline
        \acs*{dp-cdma-otfs} & $0$ dB & $0$ dB & $-2$ dB\\
        \hline
        \acs*{dd-cdma-otfs} & $-1$ dB & $+4$ dB & $-2$ dB\\
        \hline
    \end{tabular}
    \label{tab:com_res_summary_half}
    }
\end{table}

\begin{table}[tp]
    \centering
    {
    \caption{\acs*{eb/no} at which a \acs*{ber} of $10^{-4}$ is reached for \acs*{cdma-otfs} at full load relative to \acs*{otfs}}
    \begin{tabular}{|c|c|c|c|}
        \hline
        \backslashbox{Configuration}{Sequence} & Gold & Hadamard & Zadoff-Chu\\
        \hline
        \acs*{dl-cdma-otfs} & N/A & $0$ dB & $0$ dB\\
        \hline
        \acs*{dp-cdma-otfs} & N/A & $0$ dB & $0$ dB\\
        \hline
        \acs*{dd-cdma-otfs} & N/A & $+3$ dB & $0$ dB\\
        \hline
    \end{tabular}
    \label{tab:com_res_summary_full}
    }
\end{table}

\begin{table}[tp]
    \centering
    {
    \caption{\acs*{eb/no} at which the \acs*{rmse} is dominated by fractional index estimation error relative to \acs*{otfs} at full load, for a target velocity of $200$ m/s}
    \begin{tabular}{|c|c|c|c|}
        \hline
        \backslashbox{Configuration}{Sequence} & Gold & Hadamard & Zadoff-Chu\\
        \hline
        \acs*{dl-cdma-otfs} & $-1$ dB & $-2$ dB & $-1$ dB\\
        \hline
        \acs*{dp-cdma-otfs} & $-2$ dB & $-1$ dB & $-2$ dB\\
        \hline
        \acs*{dd-cdma-otfs} & $-1$ dB & $-1$ dB & $-1$ dB\\
        \hline
    \end{tabular}
    \label{tab:sen_res_summary_v_200}
    }
\end{table}

\begin{table}[tp]
    \centering
    {
    \caption{\acs*{eb/no} at which the \acs*{rmse} is dominated by fractional index estimation error relative to \acs*{otfs} at full load, for a target velocity of $110$ m/s}
    \begin{tabular}{|c|c|c|c|}
        \hline
        \backslashbox{Configuration}{Sequence} & Gold & Hadamard & Zadoff-Chu\\
        \hline
        \acs*{dl-cdma-otfs} & $-1$ dB & $-1$ dB & $-1$ dB\\
        \hline
        \acs*{dp-cdma-otfs} & $-1$ dB & $-1$ dB & $0$ dB\\
        \hline
        \acs*{dd-cdma-otfs} & $0$ dB & $-1$ dB & $-1$ dB\\
        \hline
    \end{tabular}
    \label{tab:sen_res_summary_v_110}
    }
\end{table}

Three different configurations of \ac{cdma-otfs} were introduced. The multi-user communication performance of Zadoff-Chu \ac{cdma-otfs} is similar to that of single user \ac{otfs} at an equal throughput. When fewer users are present, the multi-user throughput is diminished, and Zadoff-Chu \ac{cdma-otfs} has a lower \ac{ber} than single user \ac{otfs}. The communication performance of Zadoff-Chu sequences is similar for all three \ac{cdma-otfs} spreading configurations. Gold and Hadamard sequences do not consistently outperform single user \ac{otfs} communication. The communication performance of the different configurations relative to \ac{otfs} at the normalised half and full load are summarised in Tables \ref{tab:com_res_summary_half} and \ref{tab:com_res_summary_full}, respectively.

The three \ac{cdma-otfs} spreading configurations outperform pure \ac{otfs} sensing for all the velocities considered. Hadamard sequences lead to a superior sensing performance for \ac{dl-cdma-otfs} at high velocities, but the three sequences have a similar performance at lower velocities. Gold sequences increase the sensing performance of \ac{dp-cdma-otfs} at high velocities, whereas Zadoff-Chu sequences lead to the similar sensing performance as pure \ac{otfs} at lower velocities. The sensing performance of \ac{dd-cdma-otfs} is similar for all three sequence types at high velocities, but Gold sequences lead to an inferior sensing performance at low velocities. The sensing performance of the different configurations relative to \ac{otfs} at full load are summarised in Tables \ref{tab:sen_res_summary_v_200} and \ref{tab:sen_res_summary_v_110}, for target velocities of $200$ m/s and $110$ m/s respectively.

Following these results, Zadoff-Chu \ac{dl-cdma-otfs} and \ac{dd-cdma-otfs} are the configurations that consistently outperform pure \ac{otfs} sensing, whilst maintaining a similar communication performance at the same throughput.

The added modulation complexity of \ac{cdma-otfs} is similar to other \ac{otfs} multi-user methodologies, but the demodulation complexity of \ac{cdma-otfs} is lower than that of some other \ac{otfs} multi-user methodologies. \ac{cdma-otfs} sensing can also consistently outperform \ac{otfs} sensing whilst not requiring any additional complexity for target parameter estimation. Hence, \ac{cdma-otfs} is a computationally more attractive multi-user approach for \ac{otfs} \ac{isac} than the alternatives in the literature.

Future work will consider \ac{scma}-\ac{otfs} \ac{isac}, as there is much interest in the employment of sparse codes for multi-user systems \cite{convolutional_scma_otfs_chan_est,otfs_noma_scma_downlink_conference,otfs_scma_downlink,otfs_scma_uplink_downlink,polar_scma_iter_rec,high_dim_codebook_des_scma_dl,low_comp_energy_min_based_scma,hyrid_iter_det_dec_adpt_turbo_scma}. The sparsity of these codes is expected to reduce the sensing performance, hence methods of mitigating this effect have to be developed.

\bibliography{ref}

\begin{thebibliography}{10}
\providecommand{\url}[1]{#1}
\csname url@samestyle\endcsname
\providecommand{\newblock}{\relax}
\providecommand{\bibinfo}[2]{#2}
\providecommand{\BIBentrySTDinterwordspacing}{\spaceskip=0pt\relax}
\providecommand{\BIBentryALTinterwordstretchfactor}{4}
\providecommand{\BIBentryALTinterwordspacing}{\spaceskip=\fontdimen2\font plus
\BIBentryALTinterwordstretchfactor\fontdimen3\font minus
  \fontdimen4\font\relax}
\providecommand{\BIBforeignlanguage}[2]{{%
\expandafter\ifx\csname l@#1\endcsname\relax
\typeout{** WARNING: IEEEtran.bst: No hyphenation pattern has been}%
\typeout{** loaded for the language `#1'. Using the pattern for}%
\typeout{** the default language instead.}%
\else
\language=\csname l@#1\endcsname
\fi
#2}}
\providecommand{\BIBdecl}{\relax}
\BIBdecl

\bibitem{JRC_survey_plus_tran_design}
F.~Liu, C.~Masouros, A.~P. Petropulu, H.~Griffiths, and L.~Hanzo, ``{Joint
  Radar and Communication Design: Applications, State-of-the-Art, and the Road
  Ahead},'' \emph{IEEE Transactions on Communications}, vol.~68, no.~6, pp.
  3834--3862, 2020.

\bibitem{MU_MIMO_JCAS_and_ISAC}
F.~Liu, C.~Masouros, A.~Li, H.~Sun, and L.~Hanzo, ``{MU-MIMO Communications
  With MIMO Radar: From Co-Existence to Joint Transmission},'' \emph{IEEE
  Transactions on Wireless Communications}, vol.~17, no.~4, pp. 2755--2770,
  2018.

\bibitem{isac_human_activity_magazine}
X.~Li, Y.~Cui, J.~A. Zhang, F.~Liu, D.~Zhang, and L.~Hanzo, ``{Integrated Human
  Activity Sensing and Communications},'' \emph{IEEE Communications Magazine},
  vol.~61, no.~5, pp. 90--96, 2023.

\bibitem{isac_survey_iot_2024}
S.~Lu, F.~Liu, Y.~Li, K.~Zhang, H.~Huang, J.~Zou, X.~Li, Y.~Dong, F.~Dong,
  J.~Zhu, Y.~Xiong, W.~Yuan, Y.~Cui, and L.~Hanzo, ``{Integrated Sensing and
  Communications: Recent Advances and Ten Open Challenges},'' \emph{IEEE
  Internet of Things Journal}, vol.~11, no.~11, pp. 19\,094--19\,120, 2024.

\bibitem{otfs_first_conf}
R.~Hadani, S.~Rakib, M.~Tsatsanis, A.~Monk, A.~J. Goldsmith, A.~F. Molisch, and
  R.~Calderbank, ``{Orthogonal Time Frequency Space Modulation},'' in
  \emph{2017 IEEE Wireless Communications and Networking Conference (WCNC)},
  2017, pp. 1--6.

\bibitem{otfs_second_conf}
\BIBentryALTinterwordspacing
R.~Hadani and A.~Monk, ``{OTFS: A New Generation of Modulation Addressing the
  Challenges of 5G},'' 2018. [Online]. Available:
  \url{https://arxiv.org/abs/1802.02623}
\BIBentrySTDinterwordspacing

\bibitem{otfs_survey-ish_2021}
Z.~Wei, W.~Yuan, S.~Li, J.~Yuan, G.~Bharatula, R.~Hadani, and L.~Hanzo,
  ``{Orthogonal Time-Frequency Space Modulation: A Promising Next-Generation
  Waveform},'' \emph{IEEE Wireless Communications}, vol.~28, no.~4, pp.
  136--144, 08 2021.

\bibitem{otfs_radar_basic_conference}
P.~Raviteja, K.~T. Phan, Y.~Hong, and E.~Viterbo, ``{Orthogonal Time Frequency
  Space (OTFS) Modulation Based Radar System},'' in \emph{{2019 IEEE Radar
  Conference (RadarConf)}}, 2019, pp. 1--6.

\bibitem{OTFS_JRC_param_est}
L.~Gaudio, M.~Kobayashi, G.~Caire, and G.~Colavolpe, ``{On the Effectiveness of
  OTFS for Joint Radar Parameter Estimation and Communication},'' \emph{IEEE
  Transactions on Wireless Communications}, vol.~19, no.~9, pp. 5951--5965,
  2020.

\bibitem{otfs_rad_frac_inds_diff_est}
K.~Zhang, Z.~Li, W.~Yuan, Y.~Cai, and F.~Gao, ``{Radar sensing via OTFS
  signaling},'' \emph{China Communications}, vol.~20, no.~9, pp. 34--45, 2023.

\bibitem{otfs_com_mp_rad_frac_diff_est}
\BIBentryALTinterwordspacing
J.~Zhang, L.~Cai, and H.~Liu, ``{Integrated Sensing and Communication via
  Orthogonal Time Frequency Space Signaling with Hybrid Message Passing
  Detection and Fractional Parameter Estimation},'' \emph{Sensors}, vol.~23,
  no.~24, 2023. [Online]. Available:
  \url{https://www.mdpi.com/1424-8220/23/24/9874}
\BIBentrySTDinterwordspacing

\bibitem{otfs_isac_int_ind_mf_frac_ind_fibonacci}
Z.~Tang, Z.~Jiang, W.~Pan, and L.~Zeng, ``{The Estimation Method of Sensing
  Parameters Based on OTFS},'' \emph{IEEE Access}, vol.~11, pp.
  66\,035--66\,049, 2023.

\bibitem{otfs_chan_and_rad_param_est_letter}
S.~P. Muppaneni, S.~R. Mattu, and A.~Chockalingam, ``{Channel and Radar
  Parameter Estimation With Fractional Delay-Doppler Using OTFS},'' \emph{IEEE
  Communications Letters}, vol.~27, no.~5, pp. 1392--1396, 2023.

\bibitem{otsm_first}
T.~Thaj, E.~Viterbo, and Y.~Hong, ``{Orthogonal Time Sequency Multiplexing
  Modulation: Analysis and Low-Complexity Receiver Design},'' \emph{IEEE
  Transactions on Wireless Communications}, vol.~20, no.~12, pp. 7842--7855,
  2021.

\bibitem{otsm_letter_reduced_papr_and_pilot_power}
S.~G. Neelam and P.~R. Sahu, ``{Iterative Channel Estimation and Data Detection
  of OTSM With Superimposed Pilot Scheme and PAPR Analysis},'' \emph{IEEE
  Communications Letters}, vol.~27, no.~8, pp. 2147--2151, 2023.

\bibitem{otsm_ber_upper_bound_and_vamp_em_detector}
Z.~Sui, S.~Yan, H.~Zhang, S.~Sun, Y.~Zeng, L.-L. Yang, and L.~Hanzo,
  ``{Performance Analysis and Approximate Message Passing Detection of
  Orthogonal Time Sequency Multiplexing Modulation},'' \emph{IEEE Transactions
  on Wireless Communications}, vol.~23, no.~3, pp. 1913--1928, 2024.

\bibitem{otsm_sequency_im}
A.~Doosti-Aref, E.~Basar, and H.~Arslan, ``{Sequency Index Modulation: A Novel
  Index Modulation for Delay-Sequency Domain Waveforms},'' \emph{IEEE Wireless
  Communications Letters}, vol.~12, no.~11, pp. 1911--1915, 2023.

\bibitem{otsm_parwise_sequency_im}
A.~Doosti-Aref, C.~Masouros, E.~Basar, and H.~Arslan, ``{Pairwise Sequency
  Index Modulation With OTSM for Green and Robust Single-Carrier
  Communications},'' \emph{IEEE Wireless Communications Letters}, vol.~13,
  no.~4, pp. 1083--1087, 2024.

\bibitem{otsm_iq_imbalance_estimation_compensation_receiver}
S.~G. Neelam and P.~R. Sahu, ``{Estimation and Compensation of IQ Imbalance for
  OTSM Systems},'' \emph{IEEE Wireless Communications Letters}, vol.~11, no.~9,
  pp. 1885--1889, 2022.

\bibitem{otsm_iq_imbalance_estimation_compensation_receiver_carrier_frequency_offset_channel_estimation}
------, ``{Joint Estimation and Compensation of CFO, IQ Imbalance and Channel
  Parameters for Zero Padded OTSM Systems},'' \emph{IEEE Wireless
  Communications Letters}, vol.~12, no.~11, pp. 1871--1875, 2023.

\bibitem{otsm_iq_imbalance_estimation_compensation_transmitter_receiver_and_channel_estimation_training_sequence}
------, ``{Joint Compensation of TX/RX IQ Imbalance and Channel Parameters for
  OTSM Systems},'' \emph{IEEE Communications Letters}, vol.~27, no.~3, pp.
  976--980, 2023.

\bibitem{otsm_iq_imbalance_tx_rx_carrier_frequency_offset_dc_offset_estimation_compensation}
A.~Singh, S.~Sharma, M.~Sharma, and K.~Deka, ``{Low Complexity Deep-Decoder for
  OTSM With Hardware Impairments},'' \emph{IEEE Communications Letters},
  vol.~27, no.~12, pp. 3240--3244, 2023.

\bibitem{otfs_otsm_sc_comparison}
B.~V.~S. Reddy, C.~Velampalli, and S.~S. Das, ``{Performance Analysis of
  Multi-User OTFS, OTSM, and Single Carrier in Uplink},'' \emph{IEEE
  Transactions on Communications}, vol.~72, no.~3, pp. 1428--1443, 2024.

\bibitem{discrete_cosine_transform_otfs_papr_reduction}
N.~V. Kalpage, P.~Priya, and Y.~Hong, ``{DCT-Based OTFS With Reduced PAPR},''
  \emph{IEEE Communications Letters}, vol.~28, no.~1, pp. 158--162, 2024.

\bibitem{otfs_multi_user_DD_mat_split}
\BIBentryALTinterwordspacing
G.~D. Surabhi, R.~M. Augustine, and A.~Chockalingam, ``{Multiple Access in the
  Delay-Doppler Domain using OTFS modulation},'' 2019. [Online]. Available:
  \url{https://arxiv.org/abs/1902.03415}
\BIBentrySTDinterwordspacing

\bibitem{otfs_multi_user_DD_mat_split_guard_band_between_user_blocks}
V.~Khammammetti and S.~K. Mohammed, ``{OTFS-Based Multiple-Access in High
  Doppler and Delay Spread Wireless Channels},'' \emph{IEEE Wireless
  Communications Letters}, vol.~8, no.~2, pp. 528--531, 2019.

\bibitem{otfs_dd_matrix_partitioning_uplink_stationary_and_moving_users}
Y.~Ge, Q.~Deng, P.~C. Ching, and Z.~Ding, ``{OTFS Signaling for Uplink NOMA of
  Heterogeneous Mobility Users},'' \emph{IEEE Transactions on Communications},
  vol.~69, no.~5, pp. 3147--3161, 2021.

\bibitem{otfs_tsma}
Y.~Ma, G.~Ma, N.~Wang, Z.~Zhong, and B.~Ai, ``{OTFS-TSMA for Massive Internet
  of Things in High-Speed Railway},'' \emph{IEEE Transactions on Wireless
  Communications}, vol.~21, no.~1, pp. 519--531, 2022.

\bibitem{convolutional_scma_otfs_chan_est}
A.~Thomas, K.~Deka, P.~Raviteja, and S.~Sharma, ``{Convolutional Sparse Coding
  Based Channel Estimation for OTFS-SCMA in Uplink},'' \emph{IEEE Transactions
  on Communications}, vol.~70, no.~8, pp. 5241--5257, 2022.

\bibitem{otfs_scma_uplink_downlink}
K.~Deka, A.~Thomas, and S.~Sharma, ``{OTFS-SCMA: A Code-Domain NOMA Approach
  for Orthogonal Time Frequency Space Modulation},'' \emph{IEEE Transactions on
  Communications}, vol.~69, no.~8, pp. 5043--5058, 2021.

\bibitem{otfs_scma_downlink}
H.~Wen, W.~Yuan, Z.~Liu, and S.~Li, ``{OTFS-SCMA: A Downlink NOMA Scheme for
  Massive Connectivity in High Mobility Channels},'' \emph{IEEE Transactions on
  Wireless Communications}, vol.~22, no.~9, pp. 5770--5784, 2023.

\bibitem{otfs_ds_delay_spread_comm_only_conf}
J.~Sun, Z.~Wang, and Q.~Huang, ``{An Orthogonal Time Frequency Space Direct
  Sequence Modulation Scheme},'' in \emph{2021 IEEE International Conference on
  Communications Workshops (ICC Workshops)}, 2021, pp. 1--6.

\bibitem{otfs_ds_doppler_spread_comm_only_conf}
Y.~Cao, Z.~Qiu, and H.~Long, ``{A Sequence Spread Modulation Scheme Based on
  Orthogonal Time Frequency Space},'' in \emph{2023 IEEE 97th Vehicular
  Technology Conference (VTC2023-Spring)}, 2023, pp. 1--5.

\bibitem{otfs_noma_scma_downlink_conference}
H.~Wen, W.~Yuan, and S.~Li, ``{Downlink OTFS Non-Orthogonal Multiple Access
  Receiver Design based on Cross-Domain Detection},'' in \emph{2022 IEEE
  International Conference on Communications Workshops (ICC Workshops)}, 2022,
  pp. 928--933.

\bibitem{on_fundamental_limits_isac}
Y.~Xiong, F.~Liu, Y.~Cui, W.~Yuan, T.~X. Han, and G.~Caire, ``{On the
  Fundamental Tradeoff of Integrated Sensing and Communications Under Gaussian
  Channels},'' \emph{IEEE Transactions on Information Theory}, vol.~69, no.~9,
  pp. 5723--5751, 2023.

\bibitem{im-ofdm_isac_outperform_ofdm_isac_collection}
H.~Hawkins, C.~Xu, L.-L. Yang, and L.~Hanzo, ``{IM-OFDM ISAC Outperforms OFDM
  ISAC by Combining Multiple Sensing Observations},'' \emph{IEEE Open Journal
  of Vehicular Technology}, vol.~5, pp. 312--329, 2024.

\bibitem{otfs_ris_sagin_double_select_chan}
C.~Xu, L.~Xiang, J.~An, C.~Dong, S.~Sugiura, R.~G. Maunder, L.-L. Yang, and
  L.~Hanzo, ``{OTFS-Aided RIS-Assisted SAGIN Systems Outperform Their OFDM
  Counterparts in Doubly Selective High-Doppler Scenarios},'' \emph{IEEE
  Internet of Things Journal}, vol.~10, no.~1, pp. 682--703, 2023.

\bibitem{OTFS_and_OFDM_JRC_performance_analysis}
L.~Gaudio, M.~Kobayashi, B.~Bissinger, and G.~Caire, ``{Performance Analysis of
  Joint Radar and Communication using OFDM and OTFS},'' in \emph{2019 IEEE
  International Conference on Communications Workshops (ICC Workshops)}, 2019,
  pp. 1--6.

\bibitem{OTFS_CRB}
\BIBentryALTinterwordspacing
J.~Shi, X.~Hu, Z.~Tie, X.~Chen, W.~Liang, and Z.~Li, ``{Reliability performance
  analysis for OTFS modulation based integrated sensing and communication},''
  \emph{Digital Signal Processing}, vol. 144, p. 104280, 2024. [Online].
  Available:
  \url{https://www.sciencedirect.com/science/article/pii/S1051200423003755}
\BIBentrySTDinterwordspacing

\bibitem{polar_scma_iter_rec}
L.~Xiang, Y.~Liu, C.~Xu, R.~G. Maunder, L.-L. Yang, and L.~Hanzo, ``{Iterative
  Receiver Design for Polar-Coded SCMA Systems},'' \emph{IEEE Transactions on
  Communications}, vol.~69, no.~7, pp. 4235--4246, 2021.

\bibitem{high_dim_codebook_des_scma_dl}
L.~Li, Z.~Ma, P.~Z. Fan, and L.~Hanzo, ``{High-Dimensional Codebook Design for
  the SCMA Down Link},'' \emph{IEEE Transactions on Vehicular Technology},
  vol.~67, no.~10, pp. 10\,118--10\,122, 2018.

\bibitem{low_comp_energy_min_based_scma}
W.~Yuan, N.~Wu, C.~Yan, Y.~Li, X.~Huang, and L.~Hanzo, ``{A Low-Complexity
  Energy-Minimization-Based SCMA Detector and Its Convergence Analysis},''
  \emph{IEEE Transactions on Vehicular Technology}, vol.~67, no.~12, pp.
  12\,398--12\,403, 2018.

\bibitem{hyrid_iter_det_dec_adpt_turbo_scma}
Y.~Liu, L.~Xiang, R.~G. Maunder, L.-L. Yang, and L.~Hanzo, ``{Hybrid Iterative
  Detection and Decoding of Near-Instantaneously Adaptive Turbo-Coded Sparse
  Code Multiple Access},'' \emph{IEEE Transactions on Vehicular Technology},
  vol.~70, no.~5, pp. 4682--4692, 2021.

\end{thebibliography}

\begin{IEEEbiography}[{\includegraphics[width=1in,height=1.25in,clip,keepaspectratio]{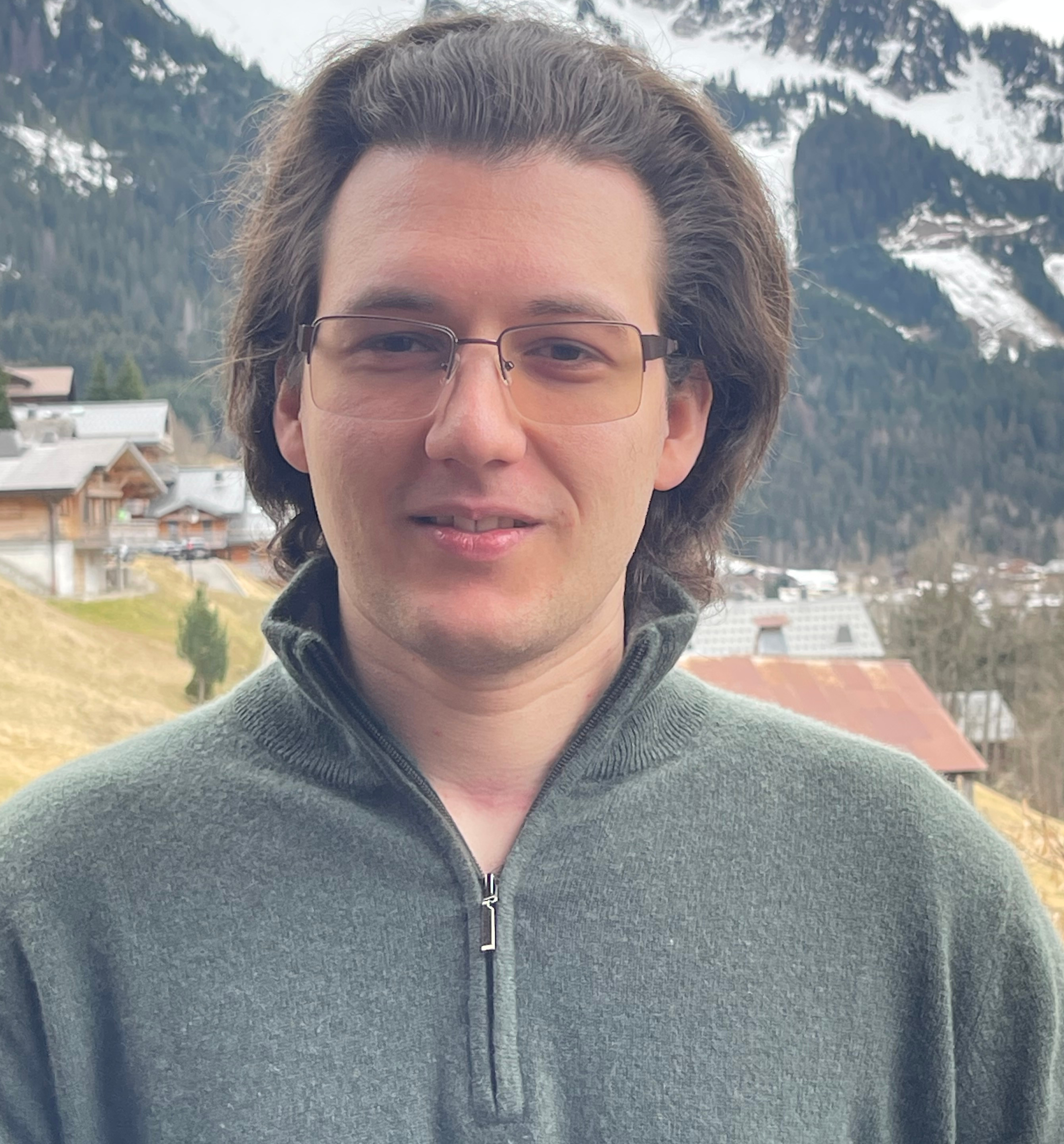}}]{Hugo Hawkins (Graduate Student Member, IEEE) } received his B.Eng. degree in Electrical and
Electronic Engineering from the University of Southampton with First Class Honours in 2021. He is currently undertaking a PhD within the Next Generation Wireless Research Group at the University of Southampton. His research interests include integrated sensing and communication.
\end{IEEEbiography}

\begin{IEEEbiography}[{\includegraphics[width=1in,height=1.25in,clip,keepaspectratio]{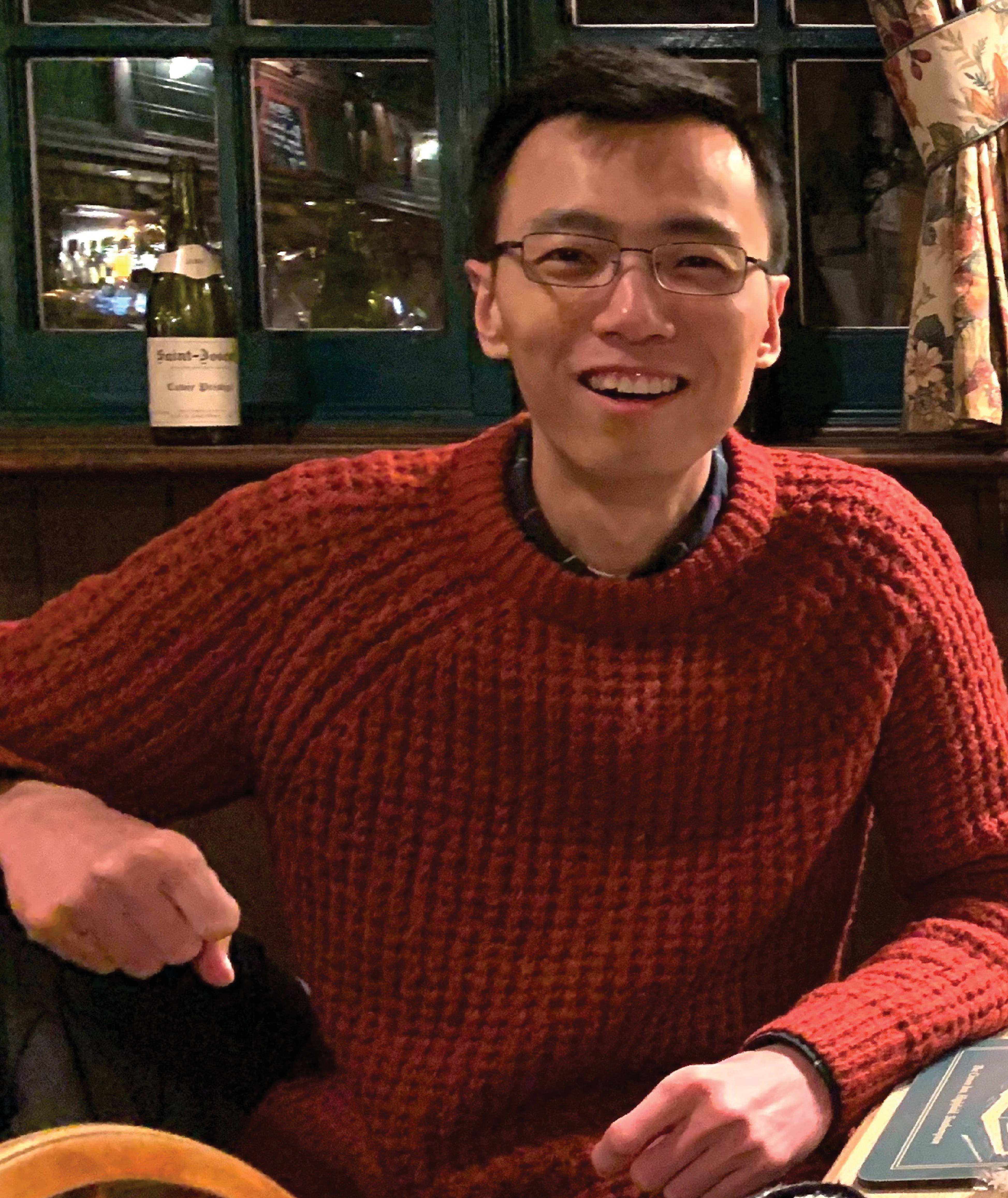}}]{Chao Xu (S'09-M'14-SM'19) } received the B.Eng. degree in telecommunications from the Beijing University of Posts and Telecommunications, Beijing, China, the B.Sc. (Eng.) degree (with First Class Hons.) in telecommunications from the Queen Mary, University of London, London, U.K., through a Sino-U.K. joint degree Program in 2008, and the M.Sc. degree (with distinction) in radio frequency communication systems and the Ph.D. degree in wireless communications from the University of Southampton, Southampton, U.K., in 2009 and 2015, respectively. He is currently a Senior Lecturer with Next Generation Wireless Research Group, University of Southampton. His research interests include index modulation, reconfigurable intelligent surfaces, noncoherent detection, and turbo detection. He was recipient of the Best M.Sc. Student in Broadband and Mobile Communication Networks by the IEEE Communications Society (United Kingdom and Republic of Ireland Chapter) in 2009. He was also the recipient of 2012 Chinese Government Award for Outstanding Self-Financed Student Abroad and 2017 Dean's Award, Faculty of Physical Sciences and Engineering, the University of Southampton. In 2023, he was awarded Marie Sklodowska-Curie Actions (MSCA) Global Postdoctoral Fellowships with the highest evaluation score of 100/100.
\end{IEEEbiography}

\begin{IEEEbiography}[{\includegraphics[width=1in,height=1.25in,clip,keepaspectratio]{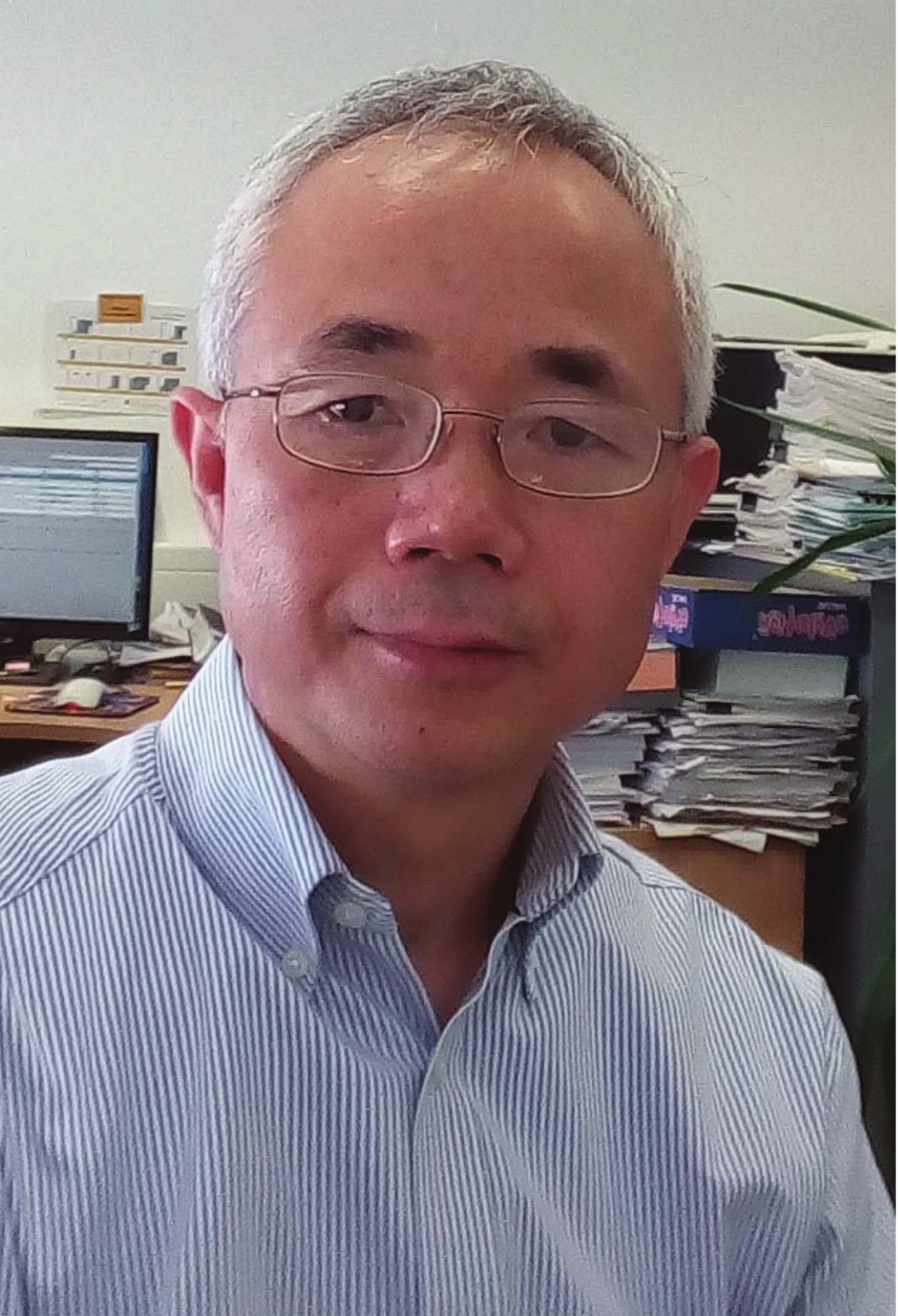}}]{Lie-Liang Yang (Fellow, IEEE) } is the professor of Wireless Communications in the School of Electronics and Computer Science at the University of Southampton, UK. He received his MEng and PhD degrees in communications and electronics from Northern (Beijing) Jiaotong University, Beijing, China in 1991 and 1997, respectively, and his BEng degree in communications engineering from Shanghai TieDao University, Shanghai, China in 1988. He has research interest in wireless communications, wireless networks and signal processing for wireless communications, as well as molecular communications and nano-networks. He has published 400+ research papers in journals and conference proceedings, authored/co-authored three books and also published several book chapters. The details about his research publications can be found at \url{https://www.ecs.soton.ac.uk/people/llyang}. He is a fellow of the IEEE, IET and AAIA, and was a distinguished lecturer of the IEEE VTS. He served as an associate editor to various journals, and is currently a senior editor to the IEEE Access and a subject editor to the Electronics Letters. He also acted different roles for organization of conferences.
\end{IEEEbiography}

\begin{IEEEbiography}[{\includegraphics[width=1in,height=1.25in,clip,keepaspectratio]{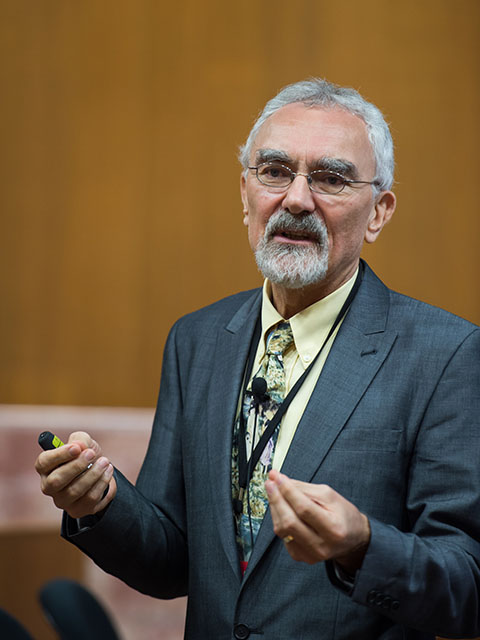}}]{LAJOS HANZO (Life Fellow, IEEE) } Lajos Hanzo (FIEEE'04) received Honorary Doctorates from the Technical University of Budapest (2009) and Edinburgh University (2015). He is a Foreign Member of the Hungarian Science-Academy, Fellow of the Royal Academy of Engineering (FREng), of the IET, of EURASIP and holds the IEEE Eric Sumner Technical Field Award. For further details please see \url{http://www-mobile.ecs.soton.ac.uk}, \url{https://en.wikipedia.org/wiki/Lajos_Hanzo}
\end{IEEEbiography}

\end{document}